\documentclass[prl,twocolumn,superscriptaddress,floatfix,preprintnumbers]{revtex4-1}
\usepackage{amsmath,amssymb,amsthm,amsfonts,mathrsfs,physics}
\usepackage{CJK}
\usepackage{color}
\usepackage{tikz,pgf}
\usetikzlibrary{shadings}
\usepackage{pgfplots,pgfplotstable}

\def \H {\mathcal{H}}
\def \A {\mathcal{A}}
\def \F {\mathcal{F}}
\def \k {\mathbf{k}}
\def \d {\hat{d}}

\begin{document}

\title{Four-Dimensional Topological Insulators with Nodal-Line Boundary States}
\author{L. B. Shao}
%\email[]{y.zhao@fkf.mpg.de}
\affiliation{National Laboratory of Solid State Microstructures and department of Physics, Nanjing University, Nanjing, 210093, China}
\affiliation{Collaborative Innovation Center of Advanced Microstructures, Nanjing University, Nanjing 210093, China}

\author{Y. X. Zhao}
\email[]{zhaoyx@nju.edu.cn}
\affiliation{National Laboratory of Solid State Microstructures and department of Physics, Nanjing University, Nanjing, 210093, China}
\affiliation{Collaborative Innovation Center of Advanced Microstructures, Nanjing University, Nanjing 210093, China}

\begin{abstract}
	Conventional topological insulators and superconductors have topologically protected nodal points on their boundaries, and the recent interests in nodal-line semimetals only concerned bulk band structures. Here, we present a novel four-dimensional topological insulator protected by an anti-unitary reflection symmetry, whose boundary band has a single $PT$-symmetric nodal line with double topological charges. Inspired by the recent experimental realization of the four-dimensional quantum Hall effect, we also propose a cold-atom system which realizes the novel topological insulator with tunable parameters as extra dimensions.
\end{abstract}

\maketitle

\textit{Introduction}
Topological insulators (TIs) and superconductors have been one of the most developed and prosperous fields in condensed matter physics during the past decade~\cite{Kane-RMP,XLQi-RMP}. We now understand that these topological phases are protected by certain symmetries, and a complete classification has been made for ten Altland-Zirnbauer (AZ) symmetry classes~\cite{AZ-classes} and for all dimensions, even beyond the physical limit of three dimensions~\cite{Schnyder-classification,Kitaev-Classification,Classification-RMP}. Historically the four-dimensional (4D) time-reversal ($T$) invariant TI played a key theoretical role in exploring topological phases~\cite{Zhang-4D-TI}, because its dimension reductions give rise to TIs in three and two dimensions~\cite{Qi-PRB-2008}. However, it has only become promising until recently to experimentally explore the topological phases in dimensions beyond three~\cite{4D-Hall-ColdAtom-Theory,4D-Hall-Cold-Atom-Exper,4D-Hall-Photonic}. Such possibilities reside in two facts. First, the topological invariants are defined in a band theory, and therefore can be well applied to artificial periodic systems other than electronic crystals, such as photonic~\cite{Haldane-photonic-crystal} and phononic~\cite{Prodan-Phonon,Kane-Phonon} crystals, and cold atoms in optical lattices~\cite{LBShao-PRL-Haldane,HZhai-PRL-Haldane%,Zak-phase-ColdAtom
}. Second, extra dimensions parametrized by momenta can be regarded as highly tunable parameters of these artificial systems, namely, that extra dimensions can be, in a sense, synthesized~\cite{Thouless-Pump,4D-Hall-ColdAtom-Theory,4D-Hall-Cold-Atom-Exper,4D-Hall-Photonic}. 

Another main focus of topological phases is the topological (semi)metals and nodal superconductors~\cite{Classification-RMP,Weyl-Dirac-Semimetal}, which mainly concerns the topological charges of band crossings~\cite{Volovik-book,ZhaoWang-Classification}. While Weyl semimetals are of fundamental interest with band crossings occuring at discrete Weyl points in the Brillouin zone (BZ)~\cite{Volovik-book,NN-NoGo,XGWan-WSM}, nodal-line semimetals, which have the band crossings forming lines in the BZ, are recently a hot topic~\cite{YuRui-Nodal-Line,Kim-PT-DNLSM,FangChen-Nodal-Line,DWZ-ColdAtom-PT,Rui-Zhao-Schnyder-DNL-Trans}. Belonging to the spacetime-inversion ($PT$) symmetric classification of topological gapless phases~\cite{Zhao-Schnyder-Wang-PT}, a nodal line can appear in generic locations in the 3D BZ with topological stability, provided $PT$ is preserved with $(PT)^2=1$~\cite{Note-Mirror-NL}.
%semimetal  nodal line protected by mirror or PT symmetry, but occurs only in bulk, we have surface nodal line

One of the significant features of TIs is that gapless modes can appear on boundaries, corresponding to nontrivial bulk topological invariants. Conventionally, the boundary gapless modes are located at isolated band-crossing points in the boundary BZ, which includes all topological phases in the periodic classification table of ten AZ symmetry classes~\cite{Schnyder-classification,Kitaev-Classification}. For instance, Weyl points of the same chirality appear on the boundary of the aforementioned 4D TI~\cite{Zhang-4D-TI,Qi-PRB-2008}. In this Letter, we present a 4D $\mathbb{Z}_2$ TI protected by an anisotropic two-fold anti-unitary spatial symmetry $RT$ with $(RT)^2=1$, where $R$ inverses the three dimensions and preserves the fourth. In contrast to boundary nodal points for conventional TIs, the crystalline TI generically has odd number of nodal-lines on the 3D boundary normal to the fourth dimension. Notably, the nodal lines have both 1D and 2D topological charges as detailed below, essentially different from ordinary nodal lines with only the 1D topological charge. We formulate the bulk $\mathbb{Z}_2$ topological invariant as the second Chern number over the half BZ, $T_{1/2}^4$, subtracted by the Chern-Simons integrals over the two $RT$-invariant boundaries of $T_{1/2}^4$. Analogous to the fact that the 4D $T$-invariant TI generalizes the 2D Chern insulator~\cite{Zhang-4D-TI,Qi-PRB-2008}, the proposed 4D crystalline TI can be regarded as a 4D generalization of the well-known 2D $T$-invariant TI~\cite{Fu-Kane-Pump}. Moreover, we discuss the general principles for realizing the topological insulator by artificial systems, and propose a cold-atom realization.

\textit{Nodal line of double topological charges--}
Let us begin by introducing the $PT$ symmetry. The spatial inversion $P$ is a unitary symmetry, which maps $\mathbf{x}$ to $-\mathbf{x}$, therefore  $\mathbf{k}$ to $-\mathbf{k}$ in momentum space, while time-reversal $T$ as an anti-unitary symmetry maps $\mathbf{k}$ to $-\mathbf{k}$ as well. Accordingly, the combination $PT$ is an anti-unitary operator that leaves any $\mathbf{k}$ invariant. For spinless fermions, it satisfies $(PT)^2=1$. The nodal-line model we would like to emerge on the boundary of a TI is
\begin{equation}
\mathcal{H}_{NL}(\mathbf{k})=\sum_{i=1}^3k_i\gamma_i+im_z\gamma_3\gamma_4. \label{Nodal-Line-Model}
\end{equation}
The $4\times 4$ hermitian matrices  $\gamma_\mu$ with $\mu=1,2\cdots,5$ satisfy the Clifford algebra $\{\gamma_\mu,\gamma_\nu\}=2\delta_{\mu\nu}$, among which $\gamma_i$ with $i=1,2,3$ are real, while $\gamma_4$ and $\gamma_5$ are purely imaginary. Thus, if $PT$ symmetry is represented as $\mathcal{PT}=\hat{ \mathcal{K}}$ with $\hat{ \mathcal{K}}$ being the complex conjugate, the model of Eq.~\eqref{Nodal-Line-Model} as a real Hamiltonian is clearly $PT$ invariant. If $m_z=0$, the Hamiltonian of Eq.~\eqref{Nodal-Line-Model} presents the real Dirac point, which can be regarded as a real generalization of the well-known Weyl points, namely the real counterpart of monopoles in the $PT$ symmetric real band theory~\cite{FangChen-Nodal-Line,Zhao-PT-Dirac}. But its monopole charge is characterized by a $\mathbb{Z}_2$ invariant $\nu^C_{\mathbb{R}}$, which is defined on a sphere $S^2$ surrounding the crossing point in the real band theory, in contrast to the $\mathbb{Z}$-valued Chern number in the complex band theory~\cite{Zhao-PT-Dirac}. The second term of Eq.~\eqref{Nodal-Line-Model} is a ``partial'' mass term, which anti-commutes with $\gamma_3$ while commutes with $\gamma_{1,2}$. Hence, when the second term of Eq.~\eqref{Nodal-Line-Model} is turned on, the four-fold degeneracy at the momentum-space origin, though, is lifted, there still exist two-fold band crossing points forming a nodal circle of $k_x^2+k_y^2=m_z^2$ in the plane with $k_z=0$~\cite{FangChen-Nodal-Line}. The fact that the spectrum cannot be fully gapped by the $PT$-symmetric term actually reflects the nontrivial topological charge $\nu_{\mathbb{R}}^C$ of the real Dirac model. It is remarkable that, while the topological charge $\nu^C_{\mathbb{R}}$ on an $S^2$ enclosing the whole nodal circle is inherited, the nodal line also has nontrivial Berry phase $\nu^B_{\mathbb{R}}$ along any loop $S^1$ surrounding it, which is quantized by $PT$ symmetry, and therefore a $\mathbb{Z}_2$ topological invariant. Thus, the Hamiltonian of Eq.~\eqref{Nodal-Line-Model} describes a $PT$-symmetric nodal-line semimetal of double topological charges.

\begin{figure}
	\includegraphics[scale=1.3]{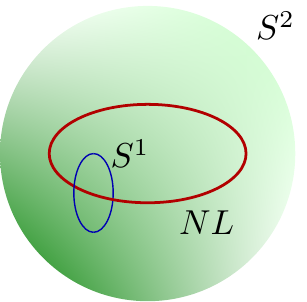}
		\caption{The nodal line with double topological charges in the boundary Brillouin zone. $S^2$ encloses the nodal line as a whole, while $S^1$ surrounds the nodal line locally in momentum space. \label{Fig-nodal-line}}
\end{figure}

\textit{The model--}
We now proceed to construct a minimal model of 4D TIs, such that the nodal line of Eq.~\eqref{Nodal-Line-Model} can correspond to its low-energy boundary theory. If there is no symmetry [except $U(1)$], the only TIs in four dimensions are second Chern insulators, for which Weyl fermions with given chirality exist on the boundary. Hence, certain symmetry is required to realize a boundary nodal line. On the other hand, we expect the bulk symmetry projected on the $3$D boundary to be $PT$ symmetry, since the nodal line is protected by $PT$ symmetry in three dimensions. Considering this condition, the symmetry we propose is an anti-unitary reflection symmetry $RT$. Here, $T$ is time-reversal symmetry, and $R$ is a unitary spatial symmetry operating as
\begin{equation}
R: ~(w,\mathbf{x}) \mapsto (w,-\mathbf{x}),~~ (k_w, \mathbf{k})\mapsto (k_w, -\mathbf{k}).
\end{equation}
in real and momentum spaces, respectively, where $\mathbf{x}$ is the three Cartesian coordinates for boundary, and $w$ is the coordinate perpendicularly towards the bulk.
Combined with time reversal, $RT$ is anti-unitary with $\{RT,i\}=0$, and operates in momentum space as
\begin{equation}
RT:~(-k_w, \mathbf{k})\mapsto (-k_w, \mathbf{k}),
\end{equation}
recalling that time reversal inverses all momentum components. In general, $RT$ can be represented in momentum space as $\mathcal{RT}=U_{RT}\hat{ \mathcal{K}}\hat{I}_w$, where $\hat{I}_w$ is the inversion of $k_w$, and $U_{RT}$ is a unitary operator. We further requires that $(\mathcal{RT})^2=1$, or equivalently $U_{RT}^T=U_{RT}$. Explicitly, $PT$ symmetry constrains the momentum-space Hamiltonian as
\begin{equation}
U_{RT}\mathcal{H}^*(\mathbf{k},k_w)U^\dagger_{RT}=\mathcal{H}(\mathbf{k},-k_w). \label{RT-H}
\end{equation}
It is clear that $\mathcal{RT}$ with $(\mathcal{RT})^2=1$  can always be converted to be $\mathcal{RT}=\hat{\mathcal{K}}\hat{I}_w$ by a unitary transformation. Therefore, we assume $\mathcal{RT}=\hat{\mathcal{K}}\hat{I}_w$ in the model construction. 

The minimal model we shall construct has eight bands, which can be inferred from the fact that the doubly charged nodal line has four bands. It is conventional for TIs that the bulk band number is a double of that of the boundary, for instance the $4$D Chern insulator has four bands with the boundary Weyl points being two-band crossings. Hence, we shall use the Clifford algebra $Cl_6$, which gives rise to a basis for $8\times 8$ matrices, as building blocks. In $8\times 8$ matrices, there are at most seven mutually anti-commuting hermitian $8\times 8$ ones, $\Gamma_a$ , namely $\{\Gamma_a,\Gamma_b\}=2\delta_{ab}$, and $\Gamma_a^\dagger=\Gamma_a$, with $a,b=0,1,\cdots,6$. Among them, following the standard convention, the first four, $\Gamma_a$ with $a=0,1,2,3$, are real, and remaining three are purely imaginary. For convenience, let $\mathbf{\Gamma}=(\Gamma_1,\Gamma_2,\Gamma_3)$, and the components be $\Gamma_i$ with $i=1,2,3$. An explicit representation of $\Gamma_a$ convenient for our purpose can be found in the Supplemental Material (SM)~\cite{Supp}.

A $RT$-symmetric massive Dirac Hamiltonian may be given by
$
\mathcal{H}^{C}_0=\mathbf{k}\cdot\mathbf{\Gamma}+k_w\Gamma_4+m\Gamma_0
$ in the continuous form. To compactify momentum space, one may replace $m$ by $m-\lambda (k^2+k^2_w)$ to get the modified Dirac model. This is so far the standard procedure to construct a TI by Dirac matrices, which however is not sufficient in our case of $Cl_6$ in four dimensions. There are other terms including $i\Gamma_i\Gamma_{5, 6}$ with $i=1,2,3$, which are real and therefore $RT$-symmetric, but still hermitian. Such terms, $i\Gamma_i\Gamma_{5, 6}$, are ``partial mass'' terms, for $i\Gamma_i\Gamma_{5, 6}$ anti-commutes with $k_i\Gamma_i$, while commutes with the rest terms in the Hamiltonian, and turn out to be essentially important for boundary in-gap modes. For simplicity, adding only $i\Gamma_3\Gamma_{5}$ into the continuous Dirac model, we can readily obtain the lattice Dirac model,
\begin{multline}
\mathcal{H} =\sum_{a=1}^{4}\sin k_i\Gamma_i+[m-(\sum_{a=1}^4\cos k_i)]\Gamma_0+it\Gamma_3\Gamma_5, \label{The-bulk-model}
\end{multline}
which is clearly $RT$ invariant with $\mathcal{RT}=\hat{ \mathcal{K}}\hat{I}_w$, and serves as a minimal model for the novel 4D TI with nodal-line boundary.

%\begin{figure}
%	\includegraphics[scale=1]{Semi-Lattice.pdf}
%	\caption{The semi-infinite lattice with a boundary perpendicular to the $w$-direction. The colored region is filled by the semi-infinite model. \label{fig-semi-infinite-lattice}}
%\end{figure}

\textit{Nodal-line boundary states}
We now open a 3D boundary for the Hamiltonian, Eq.~\eqref{The-bulk-model}, perpendicular to the $w$-direction, and discuss the in-gap boundary states of the semi-infinite system with Dirichlet boundary condition. Considering that translational symmetry is violated by the boundary in the $w$-direction, while is still preserved for the other dimensions, we introduce the ansatz $|\psi_\mathbf{k}\rangle=|\xi_{\mathbf{k}}\rangle\otimes\sum_{i=0}^{\infty}\lambda^i|i\rangle$ with $|\lambda|<1$ for boundary eigenstates, where $i$ labels the $i$th site along the $w$-direction. Solving the Schr\"{o}dinger equation for the semi-infinite system, we find that the boundary states occur in the 4D subspace corresponding to the positive eigenvalue of $i\Gamma_0\Gamma_4$ or equivalently the image of the projector $P=\frac{1}{2}(1+i\Gamma_0\Gamma_4)$, and $\lambda=m-\sum_{i=1}^3\cos k_i$ (for derivation details, see the SM)~\cite{Supp}. It is noteworthy that $i\Gamma_0\Gamma_4$ anti-commutes with $\Gamma_0$ and $\Gamma_4$, while commutes with the other $\Gamma$'s, as well as $RT$ operator $\mathcal{RT}$. Hence, the boundary low-energy effective theory is given by $\mathcal{H}_{eff}(\mathbf{k})=P\mathcal{H}P$, for the momentum $\mathbf{k}$ satisfying $|\lambda|=|m-\sum_{i=1}^3\cos k_i|<1$.
Particularly, we can focus on the eight high symmetry momenta  $\mathbf{K^\alpha}$ with $K_i^\alpha=0$ or $\pi$ and $\alpha=1,\cdots,8$. If  $|\lambda|=|m-\sum_{i=1}^3\cos K^\alpha_i|<1$, the coarse-grained theory around $\mathbf{K}^\alpha$  is explicitly given by  
\begin{equation}
\mathcal{H}^\alpha_{eff}(\mathbf{q})=\sum_{i=1}^3\eta^{\alpha,i}q_i\gamma_i+it\gamma_3\gamma_4, \label{Boundary-theory}
\end{equation} 
where $\mathbf{k}=\mathbf{K}^\alpha+\mathbf{q}$, and $\eta^{\alpha,i}$ are signs equal to $-\cos(K^\alpha_i)$. Here, we have introduced $\gamma$'s for $Cl_4$, which are specified as $\gamma_i=-P\Gamma_iP$ with $i=1,2,3$, $\gamma_4=P\Gamma_5P$ and $\gamma_5=P\Gamma_6P$, noting that $\Gamma_0$ and $\Gamma_4$ vanishes under the projection. Recall that $RT$ symmetry is projected to $PT$ in the boundary, but since $\mathcal{RT}$ commutes with the projector $P$, the low-energy effective theory of Eq.~\eqref{Boundary-theory} is clearly $PT$ invariant.

It is found that if $|m|>4$, there is no boundary state because $|\lambda|<1$ cannot be satisfied for any $\mathbf{k}$. If $2<m<4$ ($-4<m<-2$), the low-energy boundary effective theory of Eq.~\eqref{Boundary-theory} arises in the neighborhood of $\mathbf{K}=(0,0,0)$ $[(\pi,\pi,\pi)]$. If $0<m<2$ ($-2<m<0$), Eq.~\eqref{Boundary-theory} arises at three momenta, $\mathbf{K}=(\pi,0,0), (0,\pi,0), (0,0,\pi)$ [$(\pi,\pi,0), (\pi,0,\pi), (0,\pi,\pi)$]. However, in general a pair of Eq.~\eqref{Boundary-theory}'s can be gapped by $RT$-invariant terms, leaving only a single gapless low-energy theory for the last case. Thus, we find that the boundary effective theory of Eq.~\eqref{Boundary-theory} in the topological phase corresponds exactly to the nodal-line model of Eq.~\eqref{Nodal-Line-Model}, as we claimed.

\begin{figure}
	\includegraphics[scale=1]{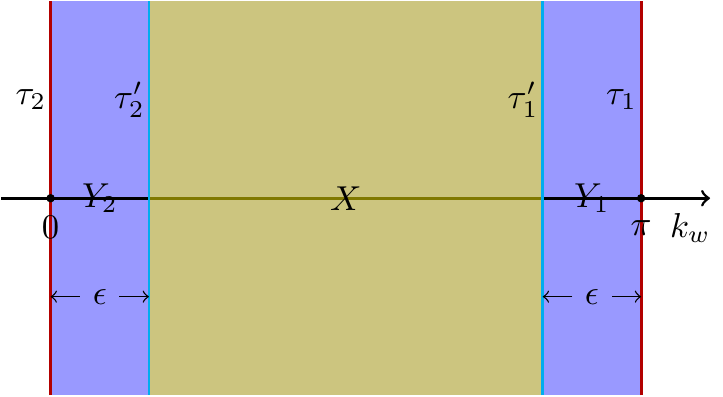}
	\caption{Two coordinate charts $X$ and $Y$ covering the half Brillouin zone. $Y$ covers the whole half BZ, but has a small singular region, which is covered by the regular chart $X$. The boundary of $X$ consists of two 3D tori, $\partial X=\tau_1'-\tau_2'$, and similarly for $Y$, $\partial Y=\tau_1-\tau_2$. Periodicity is assumed for $k_i$ with $i=1,2,3$. $Y_1$ and $Y_2$ are subspaces of $Y$, and can complement $X$ in the half Brillouin zone. \label{T4-vortex}}
\end{figure}
\textit{The bulk topological invariant--}
In momentum space $RT$ symmetry relates $(\mathbf{k},k_w)$ to $(\mathbf{k},-k_w)$. According to Eq.~\eqref{RT-H}, we can require, in a neighborhood of $(\mathbf{k},-k_w)$ and its mirror image, that
\begin{equation}
U_{RT}|\alpha,\mathbf{k},k_w\rangle^*=|\alpha,\mathbf{k},-k_w\rangle, \label{State-basis-RT}
\end{equation}
where $|\alpha,\mathbf{k},k_w\rangle$ are valence eigenstates of $\mathcal{H}(\mathbf{k},k_w)$ with $\alpha$ labeling valence bands. From Eq.~\eqref{State-basis-RT}, it is sufficient to consider only half of the BZ, $T^4_{1/2}$, as illustrated in Fig.~\ref{T4-vortex}. Particularly in two mirror-symmetric 3D tori, $\tau_1$ and $\tau_2$, namely the boundary of $T^4_{1/2}$ with $k_w=0$ and $\pi$, equation \eqref{State-basis-RT} puts constraints on eigenstates pointwisely as a boundary condition. In the nontrivial topological phase, it is impossible to find a complete basis of valence bands, which are globally well-defined in $T^4_{1/2}$, and satisfy the boundary condition of Eq.~\eqref{State-basis-RT} as well. In other words, if $|\alpha,\mathbf{k},k_w\rangle_Y$, with $Y$ being the coordinate chart covering the whole $T^4_{1/2}$, are periodic for $k_i$ with $i=1,2,3$, and satisfy Eq.~\eqref{State-basis-RT} in $\tau_1$ and $\tau_2$, there must exist singular points in the bulk of $T^4_{1/2}$ for $|\alpha,\mathbf{k},k_w\rangle_Y$ which cannot be eliminated without violating the boundary condition given by Eq.~\eqref{State-basis-RT}, while in the trivial phase, we can smooth out all singular points.  To characterize this, we first choose another coordinate chart $X$ in the bulk of $T^4_{1/2}$ (Fig.~\ref{T4-vortex}) covering all singular points of $Y$, and assume another basis of valence bands $|\alpha,\mathbf{k},k_w\rangle_X$ without singular point in $X$, which is always possible since Eq.~\eqref{State-basis-RT} puts no constraint in the bulk of $T^4_{1/2}$. Then, we can derive the transition function $t_{XY}$ from $X$ to $Y$ in the boundary of $X$, $\partial X=\tau_1'-\tau_2'$. As $t_{XY}\in U(N)$ with $N$ the valence-band number, and $\pi_3[U(N)]=\mathbb{Z}$, the transition function may has a nontrivial winding number $\mathcal{N}[t_{XY}]$. The winding number can be calculated by the Chern-Simons terms of the Berry connection on two charts restricted on $\partial X$, $\mathcal{A}_{\alpha\beta,i}^{X/Y}={}_{X/Y}\langle \alpha,k|\partial_{k_i}|\beta,k\rangle_{X/Y}$ as $\mathcal{N}=\int_{\partial X}Q_3(\mathcal{A}^X)-\int_{\partial X}Q_3(\mathcal{A}^Y)$, where $Q_3(\mathcal{A})=-\epsilon^{\mu\nu\lambda}\mathrm{tr}(\mathcal{A}_\mu\partial_\nu \mathcal{A}_\lambda+\frac{2}{3}\mathcal{A}_\mu\mathcal{A}_\nu\mathcal{A}_\lambda)/(8\pi^2) d^3k$. It can be shown that (see the SM for detailed derivations~\cite{Supp})
\begin{equation}
\mathcal{N}=\int_{T^4_{1/2}} \mathrm{ch}_2(\mathcal{F})-\int_{\tau_1}Q_3(\mathcal{A})+\int_{\tau_2} Q_3(\mathcal{A})\mod 2,\label{Invariant}
\end{equation}
where the gauge-invariant second Chern character is given by $\mathrm{ch}_2(\mathcal{F})=-\epsilon^{\mu\nu\lambda\sigma}\mathrm{tr}\mathcal{F}_{\mu\nu}\mathcal{F}_{\lambda\sigma}/(32\pi^2)d^4k$, and $\mathcal{A}$ can be chosen as $\mathcal{A}^Y$ or any basis for valence bands satisfying the boundary condition given by Eq.~\eqref{State-basis-RT}. The formula \eqref{Invariant} is a $\mathbb{Z}_2$ invariant, because a gauge transformation in either one of the Chern-Simons terms in Eq.~\eqref{Invariant} can change $\mathcal{N}$ by an even integer, noticing that $\pi_3[O(N)]=2\mathbb{Z}$ in the eight-fold periodic homotopy groups of classifying spaces in real K theory~\cite{Karoubi-book,Zhao-Schnyder-Wang-PT}. Detailed discussions on the topological invariant of the minimal model, Eq.~\eqref{The-bulk-model}, can be found in the SM~\cite{Supp}. It noteworthy that the topological invariant of Eq.~\eqref{Invariant} may be regarded as a 4D generalization of the well-known topological invariant for 2D topological insulators with a $\mathbb{Z}_2$ pump interpretation~\cite{Fu-Kane-Pump}, where the first Chern character is replaced by the second, and accordingly the Berry phases are replaced by the Chern-Simons terms.

\textit{Cold atom realization}
To simulate a 4D system by a physical system with dimension $\le 3$, certain momenta have to be simulated by tunable parameters of the physical system. For the particular Hamiltonian of Eq.~\eqref{The-bulk-model}, merely the $w$-dimension has to be faithfully simulated in real space, so that a boundary can be opened for exploring the bulk-boundary correspondence. Hence, we simulate the 4D model by a 1D lattice, and identify the three momenta $\mathbf{k}$ with highly tunable parameters of the artificial system. Then, the band structure of the in-gap boundary states can be mapped out by tuning the parameters accordingly. Based on these general considerations of simulating a high-dimensional system, the model of Eq.~\eqref{The-bulk-model}, in principle, can be realized by artificial systems, such as photonic crystals, mechanical systems, and cold atoms. We now present a simulation by cold atoms trapped in a quasi 1D optical lattice as illustrated in Fig.~\ref{simulation}. To realize the eight bands of Eq.~\eqref{The-bulk-model}, cold atoms with pseudo spin are arranged to hop on the quasi 1D tetragonal optical lattice made by four parallel chains. Observing that each term of Eq.~\eqref{The-bulk-model} is a tensor product of three Pauli matrices, we assign them to the left-right, up-down and quasi-spin spaces, respectively.
%we unitarily transform the $ \Gamma_a $ matrices to $ \Gamma_a' $ for $ a=0,1,\dots,6 $ and the Hamiltonian now is written as $ H=\sin k_w\Gamma_4'-\cos k_w\Gamma_0'+(m-\sum_{i=1}^{3}\cos k_i)\Gamma_0'+\sum_{i=1}^3\sin k_i\Gamma_i'+it\Gamma_3'\Gamma_5' $ (the detail is shown in the SM~\cite{Supp}). Since the symmetries and Clifford algebra needed are kept, the topology we derived above is the same. 
%Considering cold atoms in its ground state $ F=1 $ manifold moving in the two ladders of optical lattice vertically piled up to the bundle of four chains $ A,B,C,D $ as shown in Fig. \ref{simulation}, we take the dimension of $ w $ along the direction of the ladders and parametrize three other dimensions. The $ 8\times 8 $ matrices $ \Gamma_0, \Gamma_1,\dots, \Gamma_6 $ are the tensor product of three Pauli matrices. The first is taken as the left and right chains of both ladders $ \{\ket{l}, \ket{r}\} $, the second is the up and down ladders $\{\ket{u}, \ket{d}\} $ and the third one is the pseudo-spin $ \{\ket{\uparrow}, \ket{\downarrow}\} $ representing the internal levels $ \ket{F=1,F_z=0} $ and $ \ket{F=1,F_z=-1} $.
\begin{figure}%[h]
	\centering
	\includegraphics[scale=1.]{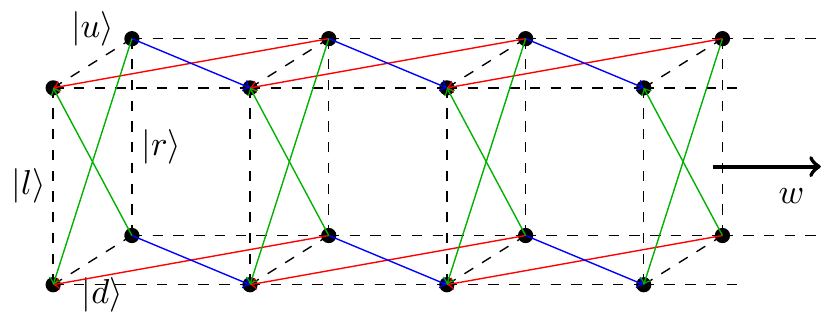}
	\caption{The quasi 1D tetragonal optical lattice. There are four sites positioned as a square in each unit cell, which are labeled as $|ur\rangle$, $|dr\rangle$, $|ul\rangle$ and $|dl\rangle$ according to their positions, respectively. All next-nearest hoppings are marked with arrows. Specifically, only green arrows involve spin-orbital interactions, and the phase difference of red and blue arrows is $\pi$.  \label{simulation}}
\end{figure}
%We set the amplitude for inter-chain nearest-neighboring (NN) hoppings as unit, and introduce all next-nearest-neighboring (NNN) hoppings by photon-assisted tunneling~\cite{2015Zhang}. 
In such a setup, the model can be well realized by the cutting-edge techniques of ultracold atom experiment. Particularly, all next-nearest-neighbor hoppings marked by colored arrows in Fig.~\ref{simulation} can be realized by photon-assisted tunnelings~\cite{2015Zhang}. The spin-orbit couplings occur merely in an inner-cell term of Eq.~\eqref{The-bulk-model} (i.e., the green hoppings in Fig.~\ref{simulation}), which can be induced by a two-photon process when a pair of incident Raman lasers, counter-propagating along the green lines in Fig.~\ref{simulation}, are resonant with the field-split internal levels~\cite{2015Zhang}. Moreover, the phases of the hoppings can be created by synthetic gauge fields~\cite{Zhu2006, 2011Atala, Miy2013}. The technical details and the realization of all terms of Eq.~\eqref{The-bulk-model} are fully addressed in the SM~\cite{Supp}.

\textit{Conclusion and Discussions} In conclusion, we have presented a 4D crystalline topological phases with nodal-line boundary states, and formulated its topological invariant. Furthermore, a minimal model has been constructed, and a cold-atom realization has been proposed. Finally, we discuss two aspects of the topological phase. First, the 2D topological charge of the nodal line is essential for the bulk boundary correspondence, not the 1D topological charge. The 2D topological charge implies that the boundary band structure with odd number of nodal lines cannot be realized by an independent 3D system, and therefore has to be connected to a higher-dimensional bulk, which follows from the index theorem of the bulk boundary correspondence~\cite{Zhao-Index}. Indeed, any 3D $PT$ symmetric semimetal satisfies the Nielsen-Ninomiya no-go theorem, namely always has even number of such nodal lines~\cite{NN-NoGo,Zhao-PT-Dirac}. Second, the topological phase is related to time-reversal symmetry with $T^2=1$, so that $(RT)^2=1$ is satisfied, provided $[R,T]=0$.  Although the requirement of $T^2=1$ is violated by electron systems with spin-orbital couplings (with $T^2=-1$), it is natural to artificial systems, such as photonic and phononic crystals, and so is to cold atoms with pseudo spin degrees of freedom.

\begin{acknowledgments}
	\textit{Acknowledgments--} We thank S. L. Zhu for discussions on the cold-atom realization. This work is supported by the startup funding of Nanjing University, China.
\end{acknowledgments}

\bibliographystyle{apsrev4-1}%{elsarticle-num}
\bibliography{4D-TI-Ref}

\begin{thebibliography}{38}%
\makeatletter
\providecommand \@ifxundefined [1]{%
 \@ifx{#1\undefined}
}%
\providecommand \@ifnum [1]{%
 \ifnum #1\expandafter \@firstoftwo
 \else \expandafter \@secondoftwo
 \fi
}%
\providecommand \@ifx [1]{%
 \ifx #1\expandafter \@firstoftwo
 \else \expandafter \@secondoftwo
 \fi
}%
\providecommand \natexlab [1]{#1}%
\providecommand \enquote  [1]{``#1''}%
\providecommand \bibnamefont  [1]{#1}%
\providecommand \bibfnamefont [1]{#1}%
\providecommand \citenamefont [1]{#1}%
\providecommand \href@noop [0]{\@secondoftwo}%
\providecommand \href [0]{\begingroup \@sanitize@url \@href}%
\providecommand \@href[1]{\@@startlink{#1}\@@href}%
\providecommand \@@href[1]{\endgroup#1\@@endlink}%
\providecommand \@sanitize@url [0]{\catcode `\\12\catcode `\$12\catcode
  `\&12\catcode `\#12\catcode `\^12\catcode `\_12\catcode `\%12\relax}%
\providecommand \@@startlink[1]{}%
\providecommand \@@endlink[0]{}%
\providecommand \url  [0]{\begingroup\@sanitize@url \@url }%
\providecommand \@url [1]{\endgroup\@href {#1}{\urlprefix }}%
\providecommand \urlprefix  [0]{URL }%
\providecommand \Eprint [0]{\href }%
\providecommand \doibase [0]{http://dx.doi.org/}%
\providecommand \selectlanguage [0]{\@gobble}%
\providecommand \bibinfo  [0]{\@secondoftwo}%
\providecommand \bibfield  [0]{\@secondoftwo}%
\providecommand \translation [1]{[#1]}%
\providecommand \BibitemOpen [0]{}%
\providecommand \bibitemStop [0]{}%
\providecommand \bibitemNoStop [0]{.\EOS\space}%
\providecommand \EOS [0]{\spacefactor3000\relax}%
\providecommand \BibitemShut  [1]{\csname bibitem#1\endcsname}%
\let\auto@bib@innerbib\@empty
%</preamble>
\bibitem [{\citenamefont {Hasan}\ and\ \citenamefont {Kane}(2010)}]{Kane-RMP}%
  \BibitemOpen
  \bibfield  {author} {\bibinfo {author} {\bibfnamefont {M.~Z.}\ \bibnamefont
  {Hasan}}\ and\ \bibinfo {author} {\bibfnamefont {C.~L.}\ \bibnamefont
  {Kane}},\ }\href {\doibase 10.1103/RevModPhys.82.3045} {\bibfield  {journal}
  {\bibinfo  {journal} {Rev. Mod. Phys.}\ }\textbf {\bibinfo {volume} {82}},\
  \bibinfo {pages} {3045} (\bibinfo {year} {2010})}\BibitemShut {NoStop}%
\bibitem [{\citenamefont {Qi}\ and\ \citenamefont {Zhang}(2011)}]{XLQi-RMP}%
  \BibitemOpen
  \bibfield  {author} {\bibinfo {author} {\bibfnamefont {X.-L.}\ \bibnamefont
  {Qi}}\ and\ \bibinfo {author} {\bibfnamefont {S.-C.}\ \bibnamefont {Zhang}},\
  }\href {\doibase 10.1103/RevModPhys.83.1057} {\bibfield  {journal} {\bibinfo
  {journal} {Rev. Mod. Phys.}\ }\textbf {\bibinfo {volume} {83}},\ \bibinfo
  {pages} {1057} (\bibinfo {year} {2011})}\BibitemShut {NoStop}%
\bibitem [{\citenamefont {Altland}\ and\ \citenamefont
  {Zirnbauer}(1997)}]{AZ-classes}%
  \BibitemOpen
  \bibfield  {author} {\bibinfo {author} {\bibfnamefont {A.}~\bibnamefont
  {Altland}}\ and\ \bibinfo {author} {\bibfnamefont {M.~R.}\ \bibnamefont
  {Zirnbauer}},\ }\href {\doibase 10.1103/PhysRevB.55.1142} {\bibfield
  {journal} {\bibinfo  {journal} {Phys. Rev. B}\ }\textbf {\bibinfo {volume}
  {55}},\ \bibinfo {pages} {1142} (\bibinfo {year} {1997})}\BibitemShut
  {NoStop}%
\bibitem [{\citenamefont {{Schnyder}}\ \emph {et~al.}(2008)\citenamefont
  {{Schnyder}}, \citenamefont {{Ryu}}, \citenamefont {{Furusaki}},\ and\
  \citenamefont {{Ludwig}}}]{Schnyder-classification}%
  \BibitemOpen
  \bibfield  {author} {\bibinfo {author} {\bibfnamefont {A.~P.}\ \bibnamefont
  {{Schnyder}}}, \bibinfo {author} {\bibfnamefont {S.}~\bibnamefont {{Ryu}}},
  \bibinfo {author} {\bibfnamefont {A.}~\bibnamefont {{Furusaki}}}, \ and\
  \bibinfo {author} {\bibfnamefont {A.~W.~W.}\ \bibnamefont {{Ludwig}}},\
  }\href {\doibase 10.1103/PhysRevB.78.195125} {\bibfield  {journal} {\bibinfo
  {journal} {\prb}\ }\textbf {\bibinfo {volume} {78}},\ \bibinfo {eid} {195125}
  (\bibinfo {year} {2008})}\BibitemShut {NoStop}%
\bibitem [{\citenamefont {Kitaev}(2009)}]{Kitaev-Classification}%
  \BibitemOpen
  \bibfield  {author} {\bibinfo {author} {\bibfnamefont {A.}~\bibnamefont
  {Kitaev}},\ }\href@noop {} {\bibfield  {journal} {\bibinfo  {journal} {AIP
  Conf. Proc.}\ }\textbf {\bibinfo {volume} {1134}},\ \bibinfo {pages} {22}
  (\bibinfo {year} {2009})}\BibitemShut {NoStop}%
\bibitem [{\citenamefont {Chiu}\ \emph {et~al.}(2016)\citenamefont {Chiu},
  \citenamefont {Teo}, \citenamefont {Schnyder},\ and\ \citenamefont
  {Ryu}}]{Classification-RMP}%
  \BibitemOpen
  \bibfield  {author} {\bibinfo {author} {\bibfnamefont {C.-K.}\ \bibnamefont
  {Chiu}}, \bibinfo {author} {\bibfnamefont {J.~C.~Y.}\ \bibnamefont {Teo}},
  \bibinfo {author} {\bibfnamefont {A.~P.}\ \bibnamefont {Schnyder}}, \ and\
  \bibinfo {author} {\bibfnamefont {S.}~\bibnamefont {Ryu}},\ }\href {\doibase
  10.1103/RevModPhys.88.035005} {\bibfield  {journal} {\bibinfo  {journal}
  {Rev. Mod. Phys.}\ }\textbf {\bibinfo {volume} {88}},\ \bibinfo {pages}
  {035005} (\bibinfo {year} {2016})}\BibitemShut {NoStop}%
\bibitem [{\citenamefont {Zhang}\ and\ \citenamefont {Hu}(2001)}]{Zhang-4D-TI}%
  \BibitemOpen
  \bibfield  {author} {\bibinfo {author} {\bibfnamefont {S.-C.}\ \bibnamefont
  {Zhang}}\ and\ \bibinfo {author} {\bibfnamefont {J.}~\bibnamefont {Hu}},\
  }\href@noop {} {\bibfield  {journal} {\bibinfo  {journal} {Science}\ }\textbf
  {\bibinfo {volume} {294}},\ \bibinfo {pages} {823} (\bibinfo {year}
  {2001})}\BibitemShut {NoStop}%
\bibitem [{\citenamefont {Qi}\ \emph {et~al.}(2008)\citenamefont {Qi},
  \citenamefont {Hughes},\ and\ \citenamefont {Zhang}}]{Qi-PRB-2008}%
  \BibitemOpen
  \bibfield  {author} {\bibinfo {author} {\bibfnamefont {X.-L.}\ \bibnamefont
  {Qi}}, \bibinfo {author} {\bibfnamefont {T.~L.}\ \bibnamefont {Hughes}}, \
  and\ \bibinfo {author} {\bibfnamefont {S.-C.}\ \bibnamefont {Zhang}},\ }\href
  {\doibase 10.1103/PhysRevB.78.195424} {\bibfield  {journal} {\bibinfo
  {journal} {Phys. Rev. B}\ }\textbf {\bibinfo {volume} {78}},\ \bibinfo
  {pages} {195424} (\bibinfo {year} {2008})}\BibitemShut {NoStop}%
\bibitem [{\citenamefont {Price}\ \emph {et~al.}(2015)\citenamefont {Price},
  \citenamefont {Zilberberg}, \citenamefont {Ozawa}, \citenamefont
  {Carusotto},\ and\ \citenamefont {Goldman}}]{4D-Hall-ColdAtom-Theory}%
  \BibitemOpen
  \bibfield  {author} {\bibinfo {author} {\bibfnamefont {H.~M.}\ \bibnamefont
  {Price}}, \bibinfo {author} {\bibfnamefont {O.}~\bibnamefont {Zilberberg}},
  \bibinfo {author} {\bibfnamefont {T.}~\bibnamefont {Ozawa}}, \bibinfo
  {author} {\bibfnamefont {I.}~\bibnamefont {Carusotto}}, \ and\ \bibinfo
  {author} {\bibfnamefont {N.}~\bibnamefont {Goldman}},\ }\href {\doibase
  10.1103/PhysRevLett.115.195303} {\bibfield  {journal} {\bibinfo  {journal}
  {Phys. Rev. Lett.}\ }\textbf {\bibinfo {volume} {115}},\ \bibinfo {pages}
  {195303} (\bibinfo {year} {2015})}\BibitemShut {NoStop}%
\bibitem [{\citenamefont {Lohse}\ \emph {et~al.}(2018)\citenamefont {Lohse},
  \citenamefont {Schweizer}, \citenamefont {Price}, \citenamefont
  {Zilberberg},\ and\ \citenamefont {Bloch}}]{4D-Hall-Cold-Atom-Exper}%
  \BibitemOpen
  \bibfield  {author} {\bibinfo {author} {\bibfnamefont {M.}~\bibnamefont
  {Lohse}}, \bibinfo {author} {\bibfnamefont {C.}~\bibnamefont {Schweizer}},
  \bibinfo {author} {\bibfnamefont {H.~M.}\ \bibnamefont {Price}}, \bibinfo
  {author} {\bibfnamefont {O.}~\bibnamefont {Zilberberg}}, \ and\ \bibinfo
  {author} {\bibfnamefont {I.}~\bibnamefont {Bloch}},\ }\href@noop {}
  {\bibfield  {journal} {\bibinfo  {journal} {Nature}\ }\textbf {\bibinfo
  {volume} {553}},\ \bibinfo {pages} {55} (\bibinfo {year} {2018})}\BibitemShut
  {NoStop}%
\bibitem [{\citenamefont {Zilberberg}\ \emph {et~al.}(2018)\citenamefont
  {Zilberberg}, \citenamefont {Huang}, \citenamefont {Guglielmon},
  \citenamefont {Wang}, \citenamefont {Chen}, \citenamefont {Kraus},\ and\
  \citenamefont {Rechtsman}}]{4D-Hall-Photonic}%
  \BibitemOpen
  \bibfield  {author} {\bibinfo {author} {\bibfnamefont {O.}~\bibnamefont
  {Zilberberg}}, \bibinfo {author} {\bibfnamefont {S.}~\bibnamefont {Huang}},
  \bibinfo {author} {\bibfnamefont {J.}~\bibnamefont {Guglielmon}}, \bibinfo
  {author} {\bibfnamefont {M.}~\bibnamefont {Wang}}, \bibinfo {author}
  {\bibfnamefont {K.~P.}\ \bibnamefont {Chen}}, \bibinfo {author}
  {\bibfnamefont {Y.~E.}\ \bibnamefont {Kraus}}, \ and\ \bibinfo {author}
  {\bibfnamefont {M.~C.}\ \bibnamefont {Rechtsman}},\ }\href@noop {} {\bibfield
   {journal} {\bibinfo  {journal} {Nature}\ }\textbf {\bibinfo {volume}
  {553}},\ \bibinfo {pages} {59} (\bibinfo {year} {2018})}\BibitemShut
  {NoStop}%
\bibitem [{\citenamefont {Haldane}\ and\ \citenamefont
  {Raghu}(2008)}]{Haldane-photonic-crystal}%
  \BibitemOpen
  \bibfield  {author} {\bibinfo {author} {\bibfnamefont {F.}~\bibnamefont
  {Haldane}}\ and\ \bibinfo {author} {\bibfnamefont {S.}~\bibnamefont
  {Raghu}},\ }\href@noop {} {\bibfield  {journal} {\bibinfo  {journal} {Phys.
  Rev. Lett.}\ }\textbf {\bibinfo {volume} {100}},\ \bibinfo {pages} {013904}
  (\bibinfo {year} {2008})}\BibitemShut {NoStop}%
\bibitem [{\citenamefont {Prodan}\ and\ \citenamefont
  {Prodan}(2009)}]{Prodan-Phonon}%
  \BibitemOpen
  \bibfield  {author} {\bibinfo {author} {\bibfnamefont {E.}~\bibnamefont
  {Prodan}}\ and\ \bibinfo {author} {\bibfnamefont {C.}~\bibnamefont
  {Prodan}},\ }\href {\doibase 10.1103/PhysRevLett.103.248101} {\bibfield
  {journal} {\bibinfo  {journal} {Phys. Rev. Lett.}\ }\textbf {\bibinfo
  {volume} {103}},\ \bibinfo {pages} {248101} (\bibinfo {year}
  {2009})}\BibitemShut {NoStop}%
\bibitem [{\citenamefont {{Kane}}\ and\ \citenamefont
  {{Lubensky}}(2014)}]{Kane-Phonon}%
  \BibitemOpen
  \bibfield  {author} {\bibinfo {author} {\bibfnamefont {C.~L.}\ \bibnamefont
  {{Kane}}}\ and\ \bibinfo {author} {\bibfnamefont {T.~C.}\ \bibnamefont
  {{Lubensky}}},\ }\href {\doibase 10.1038/nphys2835} {\bibfield  {journal}
  {\bibinfo  {journal} {Nat. Phys.}\ }\textbf {\bibinfo {volume} {10}},\
  \bibinfo {pages} {39} (\bibinfo {year} {2014})}\BibitemShut {NoStop}%
\bibitem [{\citenamefont {Shao}\ \emph {et~al.}(2008)\citenamefont {Shao},
  \citenamefont {Zhu}, \citenamefont {Sheng}, \citenamefont {Xing},\ and\
  \citenamefont {Wang}}]{LBShao-PRL-Haldane}%
  \BibitemOpen
  \bibfield  {author} {\bibinfo {author} {\bibfnamefont {L.~B.}\ \bibnamefont
  {Shao}}, \bibinfo {author} {\bibfnamefont {S.-L.}\ \bibnamefont {Zhu}},
  \bibinfo {author} {\bibfnamefont {L.}~\bibnamefont {Sheng}}, \bibinfo
  {author} {\bibfnamefont {D.~Y.}\ \bibnamefont {Xing}}, \ and\ \bibinfo
  {author} {\bibfnamefont {Z.~D.}\ \bibnamefont {Wang}},\ }\href {\doibase
  10.1103/PhysRevLett.101.246810} {\bibfield  {journal} {\bibinfo  {journal}
  {Phys. Rev. Lett.}\ }\textbf {\bibinfo {volume} {101}},\ \bibinfo {pages}
  {246810} (\bibinfo {year} {2008})}\BibitemShut {NoStop}%
\bibitem [{\citenamefont {Umucal\ifmmode \imath \else~\i \fi{}lar}\ \emph
  {et~al.}(2008)\citenamefont {Umucal\ifmmode \imath \else~\i \fi{}lar},
  \citenamefont {Zhai},\ and\ \citenamefont {Oktel}}]{HZhai-PRL-Haldane}%
  \BibitemOpen
  \bibfield  {author} {\bibinfo {author} {\bibfnamefont {R.~O.}\ \bibnamefont
  {Umucal\ifmmode \imath \else~\i \fi{}lar}}, \bibinfo {author} {\bibfnamefont
  {H.}~\bibnamefont {Zhai}}, \ and\ \bibinfo {author} {\bibfnamefont {M.~O.}\
  \bibnamefont {Oktel}},\ }\href {\doibase 10.1103/PhysRevLett.100.070402}
  {\bibfield  {journal} {\bibinfo  {journal} {Phys. Rev. Lett.}\ }\textbf
  {\bibinfo {volume} {100}},\ \bibinfo {pages} {070402} (\bibinfo {year}
  {2008})}\BibitemShut {NoStop}%
\bibitem [{\citenamefont {Thouless}(1983)}]{Thouless-Pump}%
  \BibitemOpen
  \bibfield  {author} {\bibinfo {author} {\bibfnamefont {D.~J.}\ \bibnamefont
  {Thouless}},\ }\href {\doibase 10.1103/PhysRevB.27.6083} {\bibfield
  {journal} {\bibinfo  {journal} {Phys. Rev. B}\ }\textbf {\bibinfo {volume}
  {27}},\ \bibinfo {pages} {6083} (\bibinfo {year} {1983})}\BibitemShut
  {NoStop}%
\bibitem [{\citenamefont {Armitage}\ \emph {et~al.}(2018)\citenamefont
  {Armitage}, \citenamefont {Mele},\ and\ \citenamefont
  {Vishwanath}}]{Weyl-Dirac-Semimetal}%
  \BibitemOpen
  \bibfield  {author} {\bibinfo {author} {\bibfnamefont {N.~P.}\ \bibnamefont
  {Armitage}}, \bibinfo {author} {\bibfnamefont {E.~J.}\ \bibnamefont {Mele}},
  \ and\ \bibinfo {author} {\bibfnamefont {A.}~\bibnamefont {Vishwanath}},\
  }\href {\doibase 10.1103/RevModPhys.90.015001} {\bibfield  {journal}
  {\bibinfo  {journal} {Rev. Mod. Phys.}\ }\textbf {\bibinfo {volume} {90}},\
  \bibinfo {pages} {015001} (\bibinfo {year} {2018})}\BibitemShut {NoStop}%
\bibitem [{\citenamefont {Volovik}(2003)}]{Volovik-book}%
  \BibitemOpen
  \bibfield  {author} {\bibinfo {author} {\bibfnamefont {G.~E.}\ \bibnamefont
  {Volovik}},\ }\href@noop {} {\emph {\bibinfo {title} {Universe in a helium
  droplet}}}\ (\bibinfo  {publisher} {Oxford University Press, Oxford UK},\
  \bibinfo {year} {2003})\BibitemShut {NoStop}%
\bibitem [{\citenamefont {Zhao}\ and\ \citenamefont
  {Wang}(2013)}]{ZhaoWang-Classification}%
  \BibitemOpen
  \bibfield  {author} {\bibinfo {author} {\bibfnamefont {Y.~X.}\ \bibnamefont
  {Zhao}}\ and\ \bibinfo {author} {\bibfnamefont {Z.~D.}\ \bibnamefont
  {Wang}},\ }\href {\doibase 10.1103/PhysRevLett.110.240404} {\bibfield
  {journal} {\bibinfo  {journal} {Phys. Rev. Lett.}\ }\textbf {\bibinfo
  {volume} {110}},\ \bibinfo {pages} {240404} (\bibinfo {year}
  {2013})}\BibitemShut {NoStop}%
\bibitem [{\citenamefont {Nielsen}\ and\ \citenamefont
  {Ninomiya}(1981)}]{NN-NoGo}%
  \BibitemOpen
  \bibfield  {author} {\bibinfo {author} {\bibfnamefont {H.}~\bibnamefont
  {Nielsen}}\ and\ \bibinfo {author} {\bibfnamefont {M.}~\bibnamefont
  {Ninomiya}},\ }\href {\doibase
  http://dx.doi.org/10.1016/0550-3213(81)90361-8} {\bibfield  {journal}
  {\bibinfo  {journal} {Nucl. Phys. B}\ }\textbf {\bibinfo {volume} {185}},\
  \bibinfo {pages} {20 } (\bibinfo {year} {1981})}\BibitemShut {NoStop}%
\bibitem [{\citenamefont {Wan}\ \emph {et~al.}(2011)\citenamefont {Wan},
  \citenamefont {Turner}, \citenamefont {Vishwanath},\ and\ \citenamefont
  {Savrasov}}]{XGWan-WSM}%
  \BibitemOpen
  \bibfield  {author} {\bibinfo {author} {\bibfnamefont {X.}~\bibnamefont
  {Wan}}, \bibinfo {author} {\bibfnamefont {A.~M.}\ \bibnamefont {Turner}},
  \bibinfo {author} {\bibfnamefont {A.}~\bibnamefont {Vishwanath}}, \ and\
  \bibinfo {author} {\bibfnamefont {S.~Y.}\ \bibnamefont {Savrasov}},\ }\href
  {\doibase 10.1103/PhysRevB.83.205101} {\bibfield  {journal} {\bibinfo
  {journal} {Phys. Rev. B}\ }\textbf {\bibinfo {volume} {83}},\ \bibinfo
  {pages} {205101} (\bibinfo {year} {2011})}\BibitemShut {NoStop}%
\bibitem [{\citenamefont {Yu}\ \emph {et~al.}(2015)\citenamefont {Yu},
  \citenamefont {Weng}, \citenamefont {Fang}, \citenamefont {Dai},\ and\
  \citenamefont {Hu}}]{YuRui-Nodal-Line}%
  \BibitemOpen
  \bibfield  {author} {\bibinfo {author} {\bibfnamefont {R.}~\bibnamefont
  {Yu}}, \bibinfo {author} {\bibfnamefont {H.}~\bibnamefont {Weng}}, \bibinfo
  {author} {\bibfnamefont {Z.}~\bibnamefont {Fang}}, \bibinfo {author}
  {\bibfnamefont {X.}~\bibnamefont {Dai}}, \ and\ \bibinfo {author}
  {\bibfnamefont {X.}~\bibnamefont {Hu}},\ }\href {\doibase
  10.1103/PhysRevLett.115.036807} {\bibfield  {journal} {\bibinfo  {journal}
  {Phys. Rev. Lett.}\ }\textbf {\bibinfo {volume} {115}},\ \bibinfo {pages}
  {036807} (\bibinfo {year} {2015})}\BibitemShut {NoStop}%
\bibitem [{\citenamefont {Kim}\ \emph {et~al.}(2015)\citenamefont {Kim},
  \citenamefont {Wieder}, \citenamefont {Kane},\ and\ \citenamefont
  {Rappe}}]{Kim-PT-DNLSM}%
  \BibitemOpen
  \bibfield  {author} {\bibinfo {author} {\bibfnamefont {Y.}~\bibnamefont
  {Kim}}, \bibinfo {author} {\bibfnamefont {B.~J.}\ \bibnamefont {Wieder}},
  \bibinfo {author} {\bibfnamefont {C.~L.}\ \bibnamefont {Kane}}, \ and\
  \bibinfo {author} {\bibfnamefont {A.~M.}\ \bibnamefont {Rappe}},\ }\href
  {\doibase 10.1103/PhysRevLett.115.036806} {\bibfield  {journal} {\bibinfo
  {journal} {Phys. Rev. Lett.}\ }\textbf {\bibinfo {volume} {115}},\ \bibinfo
  {pages} {036806} (\bibinfo {year} {2015})}\BibitemShut {NoStop}%
\bibitem [{\citenamefont {Zhang}\ \emph {et~al.}(2016)\citenamefont {Zhang},
  \citenamefont {Zhao}, \citenamefont {Liu}, \citenamefont {Xue}, \citenamefont
  {Zhu},\ and\ \citenamefont {Wang}}]{DWZ-ColdAtom-PT}%
  \BibitemOpen
  \bibfield  {author} {\bibinfo {author} {\bibfnamefont {D.-W.}\ \bibnamefont
  {Zhang}}, \bibinfo {author} {\bibfnamefont {Y.~X.}\ \bibnamefont {Zhao}},
  \bibinfo {author} {\bibfnamefont {R.-B.}\ \bibnamefont {Liu}}, \bibinfo
  {author} {\bibfnamefont {Z.-Y.}\ \bibnamefont {Xue}}, \bibinfo {author}
  {\bibfnamefont {S.-L.}\ \bibnamefont {Zhu}}, \ and\ \bibinfo {author}
  {\bibfnamefont {Z.~D.}\ \bibnamefont {Wang}},\ }\href {\doibase
  10.1103/PhysRevA.93.043617} {\bibfield  {journal} {\bibinfo  {journal} {Phys.
  Rev. A}\ }\textbf {\bibinfo {volume} {93}},\ \bibinfo {pages} {043617}
  (\bibinfo {year} {2016})}\BibitemShut {NoStop}%
\bibitem [{\citenamefont {{Rui}}\ \emph {et~al.}(2017)\citenamefont {{Rui}},
  \citenamefont {{Zhao}},\ and\ \citenamefont
  {{Schnyder}}}]{Rui-Zhao-Schnyder-DNL-Trans}%
  \BibitemOpen
  \bibfield  {author} {\bibinfo {author} {\bibfnamefont {W.~B.}\ \bibnamefont
  {{Rui}}}, \bibinfo {author} {\bibfnamefont {Y.~X.}\ \bibnamefont {{Zhao}}}, \
  and\ \bibinfo {author} {\bibfnamefont {A.~P.}\ \bibnamefont {{Schnyder}}},\
  }\href@noop {} {\bibfield  {journal} {\bibinfo  {journal} {ArXiv e-prints}\ }
  (\bibinfo {year} {2017})},\ \Eprint {http://arxiv.org/abs/1703.05958}
  {arXiv:1703.05958 [cond-mat.mes-hall]} \BibitemShut {NoStop}%
\bibitem [{\citenamefont {Zhao}\ \emph {et~al.}(2016)\citenamefont {Zhao},
  \citenamefont {Schnyder},\ and\ \citenamefont
  {Wang}}]{Zhao-Schnyder-Wang-PT}%
  \BibitemOpen
  \bibfield  {author} {\bibinfo {author} {\bibfnamefont {Y.~X.}\ \bibnamefont
  {Zhao}}, \bibinfo {author} {\bibfnamefont {A.~P.}\ \bibnamefont {Schnyder}},
  \ and\ \bibinfo {author} {\bibfnamefont {Z.~D.}\ \bibnamefont {Wang}},\
  }\href {\doibase 10.1103/PhysRevLett.116.156402} {\bibfield  {journal}
  {\bibinfo  {journal} {Phys. Rev. Lett.}\ }\textbf {\bibinfo {volume} {116}},\
  \bibinfo {pages} {156402} (\bibinfo {year} {2016})}\BibitemShut {NoStop}%
\bibitem [{Not()}]{Note-Mirror-NL}%
  \BibitemOpen
  \href@noop {} {}\bibinfo {note} {A nodal line on a mirror plane can also be
  protected by unitary mirror symmetry.}\BibitemShut {Stop}%
\bibitem [{\citenamefont {Fu}\ and\ \citenamefont {Kane}(2006)}]{Fu-Kane-Pump}%
  \BibitemOpen
  \bibfield  {author} {\bibinfo {author} {\bibfnamefont {L.}~\bibnamefont
  {Fu}}\ and\ \bibinfo {author} {\bibfnamefont {C.~L.}\ \bibnamefont {Kane}},\
  }\href {\doibase 10.1103/PhysRevB.74.195312} {\bibfield  {journal} {\bibinfo
  {journal} {Phys. Rev. B}\ }\textbf {\bibinfo {volume} {74}},\ \bibinfo
  {pages} {195312} (\bibinfo {year} {2006})}\BibitemShut {NoStop}%
\bibitem [{\citenamefont {Fang}\ \emph {et~al.}(2015)\citenamefont {Fang},
  \citenamefont {Chen}, \citenamefont {Kee},\ and\ \citenamefont
  {Fu}}]{FangChen-Nodal-Line}%
  \BibitemOpen
  \bibfield  {author} {\bibinfo {author} {\bibfnamefont {C.}~\bibnamefont
  {Fang}}, \bibinfo {author} {\bibfnamefont {Y.}~\bibnamefont {Chen}}, \bibinfo
  {author} {\bibfnamefont {H.-Y.}\ \bibnamefont {Kee}}, \ and\ \bibinfo
  {author} {\bibfnamefont {L.}~\bibnamefont {Fu}},\ }\href {\doibase
  10.1103/PhysRevB.92.081201} {\bibfield  {journal} {\bibinfo  {journal} {Phys.
  Rev. B}\ }\textbf {\bibinfo {volume} {92}},\ \bibinfo {pages} {081201}
  (\bibinfo {year} {2015})}\BibitemShut {NoStop}%
\bibitem [{\citenamefont {Zhao}\ and\ \citenamefont
  {Lu}(2017)}]{Zhao-PT-Dirac}%
  \BibitemOpen
  \bibfield  {author} {\bibinfo {author} {\bibfnamefont {Y.~X.}\ \bibnamefont
  {Zhao}}\ and\ \bibinfo {author} {\bibfnamefont {Y.}~\bibnamefont {Lu}},\
  }\href {\doibase 10.1103/PhysRevLett.118.056401} {\bibfield  {journal}
  {\bibinfo  {journal} {Phys. Rev. Lett.}\ }\textbf {\bibinfo {volume} {118}},\
  \bibinfo {pages} {056401} (\bibinfo {year} {2017})}\BibitemShut {NoStop}%
\bibitem [{Sup()}]{Supp}%
  \BibitemOpen
  \href@noop {} {}\bibinfo {note} {Supplemental Material}\BibitemShut {NoStop}%
\bibitem [{\citenamefont {Karoubi}(1978)}]{Karoubi-book}%
  \BibitemOpen
  \bibfield  {author} {\bibinfo {author} {\bibfnamefont {M.}~\bibnamefont
  {Karoubi}},\ }\href@noop {} {\emph {\bibinfo {title} {K-Theory: An
  Introduction}}}\ (\bibinfo  {publisher} {Springer, New York},\ \bibinfo
  {year} {1978})\BibitemShut {NoStop}%
\bibitem [{\citenamefont {{Zhang}}\ \emph {et~al.}(2015)\citenamefont
  {{Zhang}}, \citenamefont {{Cole}}, \citenamefont {{Paramekanti}},\ and\
  \citenamefont {{Trivedi}}}]{2015Zhang}%
  \BibitemOpen
  \bibfield  {author} {\bibinfo {author} {\bibfnamefont {S.}~\bibnamefont
  {{Zhang}}}, \bibinfo {author} {\bibfnamefont {W.~S.}\ \bibnamefont {{Cole}}},
  \bibinfo {author} {\bibfnamefont {A.}~\bibnamefont {{Paramekanti}}}, \ and\
  \bibinfo {author} {\bibfnamefont {N.}~\bibnamefont {{Trivedi}}},\ }\enquote
  {\bibinfo {title} {{Spin-Orbit Coupling in Optical Lattices}},}\ in\
  \href@noop {} {\emph {\bibinfo {booktitle} {Annual Review of Cold Atoms and
  Molecules - Volume 3}}},\ \bibinfo {editor} {edited by\ \bibinfo {editor}
  {\bibfnamefont {K.~W.}\ \bibnamefont {{Madison}}}}\ (\bibinfo  {publisher}
  {World Scientific Publishing Co},\ \bibinfo {year} {2015})\ pp.\ \bibinfo
  {pages} {135--179}\BibitemShut {NoStop}%
\bibitem [{\citenamefont {Zhu}\ \emph {et~al.}(2006)\citenamefont {Zhu},
  \citenamefont {Fu}, \citenamefont {Wu}, \citenamefont {Zhang},\ and\
  \citenamefont {Duan}}]{Zhu2006}%
  \BibitemOpen
  \bibfield  {author} {\bibinfo {author} {\bibfnamefont {S.-L.}\ \bibnamefont
  {Zhu}}, \bibinfo {author} {\bibfnamefont {H.}~\bibnamefont {Fu}}, \bibinfo
  {author} {\bibfnamefont {C.-J.}\ \bibnamefont {Wu}}, \bibinfo {author}
  {\bibfnamefont {S.-C.}\ \bibnamefont {Zhang}}, \ and\ \bibinfo {author}
  {\bibfnamefont {L.-M.}\ \bibnamefont {Duan}},\ }\href {\doibase
  10.1103/PhysRevLett.97.240401} {\bibfield  {journal} {\bibinfo  {journal}
  {Phys. Rev. Lett.}\ }\textbf {\bibinfo {volume} {97}},\ \bibinfo {pages}
  {240401} (\bibinfo {year} {2006})}\BibitemShut {NoStop}%
\bibitem [{\citenamefont {Aidelsburger}\ \emph {et~al.}(2011)\citenamefont
  {Aidelsburger}, \citenamefont {Atala}, \citenamefont {Nascimb\`ene},
  \citenamefont {Trotzky}, \citenamefont {Chen},\ and\ \citenamefont
  {Bloch}}]{2011Atala}%
  \BibitemOpen
  \bibfield  {author} {\bibinfo {author} {\bibfnamefont {M.}~\bibnamefont
  {Aidelsburger}}, \bibinfo {author} {\bibfnamefont {M.}~\bibnamefont {Atala}},
  \bibinfo {author} {\bibfnamefont {S.}~\bibnamefont {Nascimb\`ene}}, \bibinfo
  {author} {\bibfnamefont {S.}~\bibnamefont {Trotzky}}, \bibinfo {author}
  {\bibfnamefont {Y.-A.}\ \bibnamefont {Chen}}, \ and\ \bibinfo {author}
  {\bibfnamefont {I.}~\bibnamefont {Bloch}},\ }\href {\doibase
  10.1103/PhysRevLett.107.255301} {\bibfield  {journal} {\bibinfo  {journal}
  {Phys. Rev. Lett.}\ }\textbf {\bibinfo {volume} {107}},\ \bibinfo {pages}
  {255301} (\bibinfo {year} {2011})}\BibitemShut {NoStop}%
\bibitem [{\citenamefont {Miyake}\ \emph {et~al.}(2013)\citenamefont {Miyake},
  \citenamefont {Siviloglou}, \citenamefont {Kennedy}, \citenamefont {Burton},\
  and\ \citenamefont {Ketterle}}]{Miy2013}%
  \BibitemOpen
  \bibfield  {author} {\bibinfo {author} {\bibfnamefont {H.}~\bibnamefont
  {Miyake}}, \bibinfo {author} {\bibfnamefont {G.~A.}\ \bibnamefont
  {Siviloglou}}, \bibinfo {author} {\bibfnamefont {C.~J.}\ \bibnamefont
  {Kennedy}}, \bibinfo {author} {\bibfnamefont {W.~C.}\ \bibnamefont {Burton}},
  \ and\ \bibinfo {author} {\bibfnamefont {W.}~\bibnamefont {Ketterle}},\
  }\href {\doibase 10.1103/PhysRevLett.111.185302} {\bibfield  {journal}
  {\bibinfo  {journal} {Phys. Rev. Lett.}\ }\textbf {\bibinfo {volume} {111}},\
  \bibinfo {pages} {185302} (\bibinfo {year} {2013})}\BibitemShut {NoStop}%
\bibitem [{\citenamefont {Zhao}\ and\ \citenamefont {Wang}(2014)}]{Zhao-Index}%
  \BibitemOpen
  \bibfield  {author} {\bibinfo {author} {\bibfnamefont {Y.~X.}\ \bibnamefont
  {Zhao}}\ and\ \bibinfo {author} {\bibfnamefont {Z.~D.}\ \bibnamefont
  {Wang}},\ }\href {\doibase 10.1103/PhysRevB.89.075111} {\bibfield  {journal}
  {\bibinfo  {journal} {Phys. Rev. B}\ }\textbf {\bibinfo {volume} {89}},\
  \bibinfo {pages} {075111} (\bibinfo {year} {2014})}\BibitemShut {NoStop}%
\end{thebibliography}%

%%%%%%%%%% Merge with supplemental materials %%%%%%%%%%
\widetext
\clearpage
\begin{center}
	\textbf{\large Supplemental Materials: ``Four-Dimensional Topological Insulators with Nodal-Line Boundary States"}
\end{center}
%%%%%%%%%% Merge with supplemental materials %%%%%%%%%%
%%%%%%%%%% Prefix a "S" to all equations, figures, tables and reset the counter %%%%%%%%%%
\setcounter{equation}{0}
\setcounter{figure}{0}
\setcounter{table}{0}
\setcounter{page}{1}
\makeatletter
\renewcommand{\theequation}{S\arabic{equation}}
\renewcommand{\thefigure}{S\arabic{figure}}
\renewcommand{\bibnumfmt}[1]{[S#1]}
\renewcommand{\citenumfont}[1]{S#1}
%%%%%%%%%% Prefix a "S" to all equations, figures, tables and reset the counter %%%%%%%%%%

\section{Clifford Algebras}
For the Clifford algebra $C_{2n}$, there are $2n$ generators, namely $2^n\times 2^n$ Dirac gamma matrices $\gamma_{a}$ with $a=1,2,\cdots,2n$, satisfying
\begin{equation}
\{\gamma^a,\gamma^b\}=2\delta^{ab}.
\end{equation}
It is clear that there is the $(2n+1)$th gamma matrix anti-commuting with all $\gamma_a$ with $a=1,\cdots,2n$, which is
\begin{equation}
\gamma^{2n+1}=\pm i^n\gamma^1\gamma^2\cdots\gamma^{2n}.
\end{equation}
Note that all gamma matrices are hermitian, $\gamma_a^\dagger=\gamma_a$, and $\gamma_a^2=1$, with $a=1,2,\cdots,2n+1$. In this paper, $Cl_4$ and $ Cl_6 $ appear as building blocks of the models. Since the representation of the Clifford algebra by $2^n\times 2^n$ matrices is unique up to unitary transformations, we adopt the following convention for gamma matrices for our convenience. For $Cl_6$, the $8\times 8$ gamma matrices are given by
\begin{equation}
\begin{split}
&\Gamma_0=\sigma_1\otimes \sigma_0\otimes \sigma_0,~~~
\Gamma_1=\sigma_3\otimes\sigma_1\otimes\sigma_1,~~~ \Gamma_2=\sigma_3\otimes\sigma_1\otimes\sigma_3,~~~ \Gamma_3=\sigma_3\otimes\sigma_3\otimes \sigma_0,~~~\\ 
&\Gamma_4=\sigma_2\otimes \sigma_0\otimes \sigma_0,~~~ \Gamma_5=\sigma_3\otimes\sigma_1\otimes\sigma_2,~~~ \Gamma_6=\sigma_3\otimes\sigma_2\otimes \sigma_0,
\end{split}
\end{equation} 
where the first four $\Gamma_\mu$ with $\mu=0,1,\cdots,3$ are real, while the last three are purely imaginary. Then, we apply the projector $P=\frac{1}{2}(1+i\Gamma_0\Gamma_4)$ to get $4\times 4$ gamma matrices for $Cl_4$,
\begin{equation}
\begin{split}
&\gamma_1=\sigma_1\otimes\sigma_1,~~~ \gamma_2=\sigma_1\otimes\sigma_3,~~~ \gamma_3=\sigma_3\otimes \sigma_0,~~~\gamma_4=\sigma_1\otimes\sigma_2,~~~ \gamma_5=\sigma_2\otimes \sigma_0,
\end{split}
\end{equation} 
where $\gamma_{i}=-P\Gamma_{i}P$ with $i=1,2,3$, $\gamma_{4}=-P\Gamma_{5}P$, and $\gamma_{5}=-P\Gamma_{6}P$. Note that $P\Gamma_{0,4}P=0$.

\section{The boundary effective theory}
To open a boundary perpendicular to the $w$-direction, we first apply the inverse Fourier transform for $k_w$ to get the first quantized Hamiltonian with the $w$-dimension in real space,
\begin{equation}
\H=\sum_{i=1}^3\sin k_i\Gamma_i+\frac{1}{2i}(S-S^\dagger)\Gamma_4+\left[m-\sum_{i=1}^3\cos k_i-\frac{1}{2}(S+S^\dagger)\right]\Gamma_0+it\Gamma_3\Gamma_5
\end{equation}
where $S$ is the translation operator along the $w$-direction, and $S^\dagger$ is that along the $-w$-direction. So, $S|i\rangle=|i+1\rangle$ and $S^\dagger|i\rangle=|i-1\rangle$ with integer $i$ numbering lattice site of the $w$-dimension, and accordingly the matrices are 
\begin{equation}
S=\begin{pmatrix}
\ddots & \vdots & \vdots & \vdots & \vdots & \scalebox{-1}[1]{$\ddots$}\\
\ddots&0 & 0 & 0 & 0  &\cdots \\
\cdots&1 & 0 & 0 & 0  &\cdots\\
\cdots&0 & 1 & 0 & 0  &\cdots\\
\cdots&0 & 0 & 1 & 0  & \cdots\\
\scalebox{-1}[1]{$\ddots$} & \vdots & \vdots & \vdots & \ddots & \ddots
\end{pmatrix},\quad S^\dagger=\begin{pmatrix}
\ddots & \ddots & \vdots & \vdots & \vdots & \scalebox{-1}[1]{$\ddots$}\\
\cdots&0 & 1 & 0 & 0  &\cdots \\
\cdots&0 & 0 & 1 & 0  &\cdots\\
\cdots&0 & 0 & 0 & 1  &\cdots\\
\cdots&0 & 0 & 0 & 0  & \ddots\\
\scalebox{-1}[1]{$\ddots$} & \vdots & \vdots & \vdots & \vdots & \ddots
\end{pmatrix}.
\end{equation}
If the boundary is opened with the system being on the non-negative part of the $w$-axis, the corresponding Hamiltonian is given by
\begin{equation}
\widehat{\H}=\sum_{i=1}^3\sin k_i\Gamma_i+\frac{1}{2i}(\widehat{S}-\widehat{S}^\dagger)\Gamma_4+\left[m-\sum_{i=1}^3\cos k_i-\frac{1}{2}(\widehat{S}+\widehat{S}^\dagger)\right]\Gamma_0+it\Gamma_3\Gamma_5, \label{Half-H}
\end{equation}
where still $\widehat{S}|i\rangle=|i+1\rangle$ with $i=0,1,2,\cdots$, but $\widehat{S}^\dagger|0\rangle=0$ and $\widehat{S}^\dagger|i\rangle=|i-1\rangle$ with $i\ge 1$. Explicitly,
\begin{equation}
\widehat{S}=\begin{pmatrix}
0 & 0 & 0 & 0  &\cdots \\
1 & 0 & 0 & 0  &\cdots\\
0 & 1 & 0 & 0  &\cdots\\
0 & 0 & 1 & 0  & \cdots\\
\vdots & \vdots & \vdots & \ddots & \ddots
\end{pmatrix},\quad 
\widehat{S}^\dagger=\begin{pmatrix}
0 & 1 & 0 & 0  &\cdots \\
0 & 0 & 1 & 0  &\cdots\\
0 & 0 & 0 & 1  &\cdots\\
0 & 0 & 0 & 0  & \ddots\\
\vdots & \vdots & \vdots & \vdots & \ddots
\end{pmatrix}.
\end{equation}
We now solve the Schr\"{o}dinger equation for Eq.~\eqref{Half-H} by substituting the ansatz $|\psi_\mathbf{k}\rangle=|\xi_{\mathbf{k}}\rangle\otimes\sum_{i=0}^{\infty}\lambda^i|i\rangle$ with $|\lambda|<1$ for boundary eigenstates. Inside the bulk with $|i\ge 1\rangle$, the Schr\"{o}dinger equation gives 
\begin{align}
\left[\sum_{i}\sin k_i\Gamma_{i} +\frac{1}{2i}(\lambda-\lambda^{-1})\Gamma_4 +\left(m-\sum_{i}\cos k_i-\frac{1}{2}(\lambda+\lambda^{-1})\right) \Gamma_0+it\Gamma_3\Gamma_5 \right]\ket{\xi}=\mathcal{E}\ket{\xi}\label{bulk},
\end{align}
while at $|i=0\rangle$
\begin{align}
\left[\sum_{i}\sin k_i\Gamma_{i} +\frac{1}{2i}\lambda\Gamma_4 +\left(m-\sum_{i}\cos k_i-\frac{1}{2}\lambda\right) \Gamma_0+it\Gamma_3\Gamma_5 \right]\ket{\xi}=\mathcal{E}\ket{\xi}.\label{boundary}
\end{align}
The difference of Eqs.~\eqref{bulk} and \eqref{boundary} gives
\begin{equation}
i\Gamma_0\Gamma_4\ket{\xi}=\ket{\xi}. \label{Sub-space-condition}
\end{equation}
This equation implies the boundary states occur merely in the 4D subspace with positive eigenvalue $1$ of $i\Gamma_0\Gamma_4$, which corresponds to the projector
\begin{equation}
P=\frac{1}{2}(1+i\Gamma_0\Gamma_4).
\end{equation}
Applying the projector to Eq.~\eqref{boundary}, we have
\begin{equation}
\left(\sum_{i}\sin k_i\Gamma_{i}+it\Gamma_3\Gamma_5 \right)\ket{\xi}=\mathcal{E}\ket{\xi}, \label{Boundary-equation}
\end{equation}
The difference of Eqs.~\eqref{Boundary-equation} and \eqref{boundary}, together with Eq.~\eqref{Sub-space-condition}, gives
\begin{equation}
\lambda=m-\sum_{i}\cos k_i. \label{lambda}
\end{equation}
The effective Hamiltonian for boundary states is just 
\begin{equation}
\mathcal{H}_{eff}(\mathbf{k})=P\mathcal{H}(\k)P
\end{equation}
for regions in $\mathbf{k}$ space satisfying Eq.~\eqref{lambda}.

%\newpage
\section{The Topological Invariant}
In this section, we first give a detailed derivation of the topological invariant in the main text, and then explicitly calculate the topological invariant of the Dirac model.
\subsection{The formula}
For any region with a well-defined Berry connection, the second Chern character is the total derivative of the Chern-Simons form, and therefore we can apply Stokes' theorem to obtain the following identities,
\begin{eqnarray}
\int_X \mathrm{ch}_2(\F)&=&\int_{\partial X}Q_3(\A^X),\\ \int_{Y_1}\mathrm{ch}_2(\F)&=&\int_{\tau_1}Q_3(\A^Y)-\int_{\tau_1'}Q_3(\A^Y),\\
\int_{Y_2} \mathrm{ch}_2(\F)&=&\int_{\tau'_2}Q_3(\A^Y)-\int_{\tau_2}Q_3(\A^Y).
\end{eqnarray}

With these identities, we proceed that 
\begin{equation}
\begin{split}
\mathcal{N} &=\int_{\partial X}Q_3(\A^X)-\int_{\partial_X}Q_3(\A^Y)\\
&=\int_X \mathrm{ch}_2(\F)-\int_{\tau_1'}Q_3(\A^Y)+\int_{\tau_2'}Q_3(\A^Y)\\
&=\int_X\mathrm{ch}_2(\F)+\int_{Y_1+Y_2}\mathrm{ch}_2(\F)-\int_{\tau_1}Q_3(\A^Y)+\int_{\tau_2}Q_3(\A^Y)\\
&=\int_{T^4_{1/2}}\mathrm{ch}_2(\F)-\int_{\tau_1}Q_3(\A^Y)+\int_{\tau_2}Q_3(\A^Y)\\
&=\int_{T^4_{1/2}}\mathrm{ch}_2(\F)-\int_{\tau_1}Q_3(\A)+\int_{\tau_2}Q_3(\A)\mod 2.
\end{split}
\end{equation}
In the last equality, the Chern-Simons forms can be derived from any valence basis satisfying the boundary condition on $\tau_1$ and $\tau_2$ imposed by $RT$ symmetry, and  we have used the fact that a gauge transformation preserving the boundary condition can change either Chern-Simons term by an even number, namely, that only the expression in the last line is gauge invariant. 

\subsection{The second Chern number}
Since the topological invariant is preserved under adiabatic deformations of the Hamiltonian without closing the energy gap, let us set $t=0$ throughout this section for technical simplification. We first calculate the second Chern number in the half BZ, $T^4_{1/2}$. For this purpose, we need to derive a complete set of globally well-defined wave functions in the BZ for valence bands, of which the existence is clear because $RT$ symmetry makes the Chern number in the whole BZ vanishing.
Introducing 
\begin{eqnarray}
R_0(k) &=& m-\sum_{i=1}^3\cos k_i-\cos k_w,\\
R_i(k)&=&\sin k_i,\quad i=1,2,3,\\
R_4(k) &=& \sin k_w,
\end{eqnarray}
we write the Hamiltonian in the concise form,
\begin{equation}
\mathcal{H}(k)=\sum_{a=0}^4R_{a}(k)\Gamma^a.
\end{equation}
Applying the unitary transformation,
\begin{equation}
U=e^{\frac{\pi}{4}\Gamma_0\Gamma_4} e^{\frac{\pi i}{4}\Gamma_{0}}e^{-\frac{\pi}{4}\Gamma_4\Gamma_6},
\end{equation}
we have
\begin{equation}
\tilde{\Gamma}_\mu=U\Gamma_\mu U^\dagger
\end{equation}

\begin{equation}
\tilde{\Gamma}_0=\begin{pmatrix}
0 & i1_4\\
-i 1_4 & 0
\end{pmatrix},\quad
\tilde{\Gamma}_i=\begin{pmatrix}
0 & \tilde{\gamma}_i\\
\tilde{\gamma}_i & 0
\end{pmatrix},\quad  \tilde{\Gamma}_4=\begin{pmatrix}
0 & \tilde{\gamma}_4\\
\tilde{\gamma}_4 & 0
\end{pmatrix},
\end{equation}
where 
\begin{equation}
\tilde{\gamma}_1=\sigma_1\otimes\sigma_1,\quad \tilde{\gamma}_2=\sigma_1\otimes\sigma_3,\quad \tilde{\gamma}_3=\sigma_3\otimes\sigma_0,\quad \tilde{\gamma}_4=\sigma_2\otimes\sigma_0,
\end{equation} 
and thereby
\begin{equation}
U\mathcal{H}U^\dagger=\begin{pmatrix}
0 & Q^\dagger\\
Q & 0
\end{pmatrix}
\end{equation}
with
\begin{equation}
Q=-i R_0+R_j\tilde{\gamma}^j.
\end{equation}
We can renormalize $Q$ to be a unitary matrix,
\begin{equation}
\hat{Q}=-i \hat{R}_0+\hat{R}_j\tilde{\gamma}^j,
\end{equation}
where $\hat{R}_\mu=R_\mu/R$ with $R=\sqrt{\sum_\mu R_\mu^2}$ and $\mu=0,1,2,3$. It is noteworthy that 
\begin{equation}
\hat{Q}^T(\k,k_w)=\hat{Q}(\k,-k_w),
\end{equation}
as required by $RT$ symmetry.
Accordingly, the valence eigenstates are given by
\begin{equation}
\begin{split}
|\alpha,k\rangle&=\frac{1}{\sqrt{2}}U^\dagger\begin{pmatrix}
-\chi_\alpha(k)\\
\hat{Q}(k)\chi_{\alpha}(k)
\end{pmatrix}%\\
%&=\frac{1}{2}\begin{pmatrix}
%(i\hat{Q}-1)\xi_\alpha\\
%(\hat{Q}-i)\xi_\alpha
%\end{pmatrix},
\end{split}\label{Wave-functions}
\end{equation}
where $\chi_\alpha(k)$ is an orthonormal basis of the 4D Hilbert space for valence bands at each $k$, namely
\begin{equation}
\chi^\dagger_\alpha(k)\chi_\beta(k)=\delta_{\alpha\beta}.
\end{equation}
For convenience, we choose
\begin{equation}
\chi_1=\begin{pmatrix}
1\\
0\\
0\\
0
\end{pmatrix},\quad 
\chi_2=\begin{pmatrix}
0\\
1\\
0\\
0
\end{pmatrix}\quad 
\chi_3=\begin{pmatrix}
0\\
0\\
1\\
0
\end{pmatrix},\quad 
\chi_4=\begin{pmatrix}
0\\
0\\
0\\
1
\end{pmatrix}. \label{Chi-vectors}
\end{equation}

The globally well-defined wave functions of Eq.~\eqref{Wave-functions} with Eq.~\eqref{Chi-vectors} for valence bands give the concise expression for the Berry connection,
\begin{equation}
\A^\mu=\frac{1}{2}\hat{Q}^\dagger \partial_\mu \hat{Q}. \label{Berry-connection}
\end{equation}
Substituting Eq.~\eqref{Berry-connection} into the formula for the Berry curvature,
\begin{equation}
\F_{\mu\nu}=\partial_\mu\A_{\nu}-\partial_\nu\A_{\mu}+[\A_\mu,\A_\nu]
\end{equation}
we find
\begin{equation}
\F_{\mu\nu}=\frac{1}{4}(\partial_\mu \hat{Q}^\dagger \partial_\nu\hat{Q}-\partial_\nu \hat{Q}^\dagger \partial_\mu\hat{Q}),
\end{equation}
and thereby
\begin{equation}
\mathrm{ch}_2=-\frac{1}{128\pi^2}\epsilon^{\mu\nu\lambda\sigma}\mathrm{tr}\partial_\mu \hat{Q}^\dagger \partial_\nu\hat{Q}\partial_\lambda \hat{Q}^\dagger \partial_\sigma\hat{Q},
\end{equation}
which turns out to be vanished after tedious derivations. Thus, the topological invariant for this model is determined merely by the boundary Chern-Simons terms, derived from valence wave functions with the boundary condition given by $RT$ symmetry.

\subsection{The representation of $RT$ symmetry}
We now discuss how the $RT$ operator $\mathcal{RT}=U_{RT}\hat{\mathcal{K}}\hat{I}_w$ is represented in the valence band structure. First, we have the identities,
\begin{equation}
U_{RT} U_{RT}^\dagger=1, \quad U_{RT}=U_{RT}^T.
\end{equation}
The symmetry operator $\mathcal{RT}=U_{RT}\hat{\mathcal{K}}\hat{I}_w$ gives the constraint on the Hamiltonian,
\begin{equation}
U_{RT} \mathcal{H}^*(\mathbf{k},k_w)U^\dagger_{RT}=\mathcal{H}(\mathbf{k},-k_w),
\end{equation}
which implies $RT$ symmetry is represented by the wave functions as

\begin{equation}
U_{RT}|\alpha,\mathbf{k},k_w\rangle^*=|\beta,\mathbf{k},-k_w\rangle \mathcal{U}_{\beta\alpha}(\mathbf{k},-k_w).
\end{equation}
\begin{equation}
\mathcal{U}_{\alpha\beta}(\mathbf{k},k_w)\mathcal{U}_{\beta\gamma}^*(\mathbf{k},-k_w)=\delta_{\alpha\gamma}, \quad \mathcal{U}^T_{\alpha\beta}(\mathbf{k},k_w)=\mathcal{U}_{\alpha\beta}(\mathbf{k},-k_w).
\end{equation}
The topological obstruction in the topological phase can be essentially stated as that there exist no globally well-defined valence basis that can simultaneously  ``diagonalize'' $\mathcal{H}(k)$ and $\mathcal{RT}$ in the whole BZ. In other words, if we require that 
\begin{equation}
\mathcal{U}(\k,0)=U_{RT}=\mathcal{U}(\k,\pi), \label{Boundary-condition}
\end{equation}
which is just the boundary condition of Eq.~(7) in the main text, the valence basis must be singular at some points in the bulk of the half BZ, $T^4_{1/2}$. On the other hand, if the basis is well-defined in the whole BZ, the boundary condition of Eq.~\eqref{Boundary-condition} must be violated.

In particular, the wave functions of Eq.~\eqref{Wave-functions} are clearly well defined in the whole BZ, provided $\chi_\alpha$ are well-defined periodic functions, and accordingly give
\begin{equation}
\mathcal{U}_{\alpha\beta}(\k,k_w)=i\chi^\dagger_\alpha(\k,k_w)\hat{Q}^\dagger(\k,k_w)\chi^*_\beta(\k,-k_w).
\end{equation}
In this case, $U_{RT}$ is just the identity matrix, and one can verify that there are no $\chi_\alpha(k)$ that can transform $\hat{Q}^\dagger$ into identity matrix in the two boundaries with $k_w=0$ and $\pi$ at the same time in the topological phases.

%In two symmetric sub BZ's with $k_w=0$ and $\pi$, respectively,
%\begin{eqnarray}
%Q(0)&=&-i(m-1-\sum_{i=1}^3\cos k_i)+\sum_{i=1}^3\sin k_i\tilde{\gamma}^i,\\
%Q(\pi)&=&-i(m+1-\sum_{i=1}^3\cos k_i)+\sum_{i=1}^3\sin k_i\tilde{\gamma}^i.
%\end{eqnarray}

%We now discuss the diagonalization of $\hat{Q}^{\dagger}$. 
%\begin{equation}
%\hat{Q}^\dagger=i\gamma_4 \gamma^\mu\hat{d}_\mu,
%\end{equation}
%where $\gamma^4=\tilde{\gamma}_4$, and $\gamma^i=-i\gamma_4\tilde{\gamma}^i$ with $i=1,2,3$.
%Then introducing
%\begin{equation}
%\sigma^{\alpha\beta}=\frac{i}{4}[\gamma^\alpha,\gamma^\beta],
%\end{equation}
%we can apply the spinor representation of $SO(4)$,
%\begin{equation}
%e^{i\omega_{\alpha\beta}\sigma^{\alpha\beta}}i\gamma_4\gamma^\mu e^{-i\omega_{\alpha\beta}\sigma^{\alpha\beta}}=[e^{i\omega_{\alpha\beta}\mathcal{J}^{\alpha\beta}}]_{\mu\nu} i\gamma_4\gamma^\nu.
%\end{equation}
%Accordingly, 
%\begin{equation}
%\hat{d_\mu}\rightarrow [e^{-i\omega\cdot\mathcal{J}}]_{\mu\nu}\hat{d}_\nu.
%\end{equation}

%\section{The second Chern number}

\subsection{The boundary Chern-Simons terms}

The Hamiltonian restricted on the two boundaries $\tau_1$ and $\tau_2$ is
\begin{equation}
\H(\mathbf{k},\pi/0)=\sum_{i=1}^3\sin k_i\Gamma_{i}+\left (m\pm 1-\sum_{i=1}^{3}\cos k_i\right)\Gamma_4,
\end{equation}
In principle, there exists a global basis in $\tau_1$ ($\tau_2$) with the boundary condition for valence bands, $|\alpha,\k\rangle$
with $\alpha=1,2,3,4$, for $\H(\k,\pi)$ ($\H(\k,0)$), because there is no 3D topological insulators protected by $PT$ symmetry with $(PT)^2=1$. Note that $RT$ restricted on either $\tau_1$ or $\tau_2$ is just $PT$. However, to derive the global wave function is quite technically involved.

To warm up, we diagonalize the 1D Hamiltonian,
\begin{equation}
\H(k)=\cos k~\sigma_3\otimes \sigma_0+\sin k~\sigma_1\otimes\sigma_1.
\end{equation}
It is easy to obtain the two sets of valence eigenstates,
\begin{equation}
\psi_1(k)=\begin{pmatrix}
\sin k \\0\\0\\-1-\cos k
\end{pmatrix},\quad \psi_2(k)=\begin{pmatrix}
0\\ \sin k\\-1-\cos k\\0
\end{pmatrix},\quad k\in[-\frac{\pi}{2},\frac{\pi}{2}]
\end{equation}
and
\begin{equation}
\tilde{\psi}_1(k)=\begin{pmatrix}
1-\cos k \\0\\0\\-\sin k
\end{pmatrix},\quad \tilde{\psi}_2(k)=\begin{pmatrix}
0\\ 1-\cos k\\-\sin k\\0
\end{pmatrix},\quad k\in[\frac{\pi}{2},\frac{3\pi}{2}],
\end{equation}
where $\psi_{1,2}$ and $\tilde{\psi}_{1,2}$ have the singular points $k=\pi$ and $k=0$, respectively. To achieve a global basis, we shall glue them at $k=-\frac{\pi}{2}$ and $\frac{\pi}{2}$ periodically and continuously.
Through they are equal at $k=\pi/2$, they are not smoothly connected at ${3\pi}/{2}$ and $-\pi/2$. The following recombination realizes the continuity,
\begin{eqnarray}
(\psi_1(k)+\psi_2(k))/\sqrt{2},\quad (\psi_1(k)-\psi_2(k))/\sqrt{2},\quad && k\in[-\pi/2,\pi/2]\\
\cos(k/2)\tilde{\psi}_1(k)+\sin(k/2)\tilde{\psi}_2(k),\quad \sin(k/2)\tilde{\psi}_1(k)-\cos(k/2)\tilde{\psi}_2(k),\quad && k\in[-\pi/2,\pi/2]
\end{eqnarray}
After further simplification, we arrive at the globally well-defined valence orthonormal basis,
\begin{equation}
|1,k\rangle=\frac{1}{2}\begin{pmatrix}
\sin k\\ 1-\cos k \\ -\sin k\\-1-\cos k
\end{pmatrix},\quad |2,k\rangle=\frac{1}{2}\begin{pmatrix}
1-\cos k\\ -\sin k \\ 1+\cos k\\-\sin k
\end{pmatrix}
\end{equation}
which can also be derived as
\begin{equation}
|1,k\rangle=\frac{1}{2}(\psi_1+\tilde{\psi}_2),\quad 
|2,k\rangle=\frac{1}{2}(\tilde{\psi}_1-\psi_2).
\end{equation}

With the insight from the above simple model, we now return to solving the Hamiltonian. First, we flatten the Hamiltonian through dividing it by  $d=R(\k,\pi/0)$,
\begin{equation}
\begin{split}
\tilde{\H} &=\H(\k,\pi/0)/d\\
&=\hat{d}_0\Gamma_0+\hat{d}_i\Gamma^i\\
&=\begin{pmatrix}
\hat{d}_0 & \hat{\Delta}\\
\hat{\Delta} & -\hat{d}_0
\end{pmatrix},
\end{split}
\end{equation}
where $\sum_{\mu=0}^3 \d_\mu^2=1$, and $\hat{\Delta}$ is a real symmetric matrix,
\begin{equation}
\hat{\Delta}=\begin{pmatrix}
\hat{d}_3 & 0 & \hat{d}_2 & \hat{d}_1\\
0  & \hat{d}_3 & \hat{d}_1 & -\hat{d}_2\\
\hat{d}_2 & \hat{d}_1 & -\hat{d}_3 & 0\\
\hat{d}_1 & -\hat{d}_2 & 0 & -\hat{d}_3
\end{pmatrix}.
\end{equation}
Then, 
\begin{equation}
\psi_1(\xi_1)=\begin{pmatrix}
\hat{\Delta}\xi_1\\
-(1+\hat{d}_0)\xi_1
\end{pmatrix},\quad 
\psi_2(\xi_2)=\begin{pmatrix}
(1-\hat{d}_0)\xi_2\\
-\hat{\Delta}\xi_2
\end{pmatrix}
\end{equation}
are two forms of eigenstates. $\psi_{1,2}$ is singular if and only if $\hat{d}_0=\mp 1$, and therefore $\psi_{1}$ and $\psi_{2}$ cannot both be singular at the same $\k$. To construct globally well-defined wave functions, we combine them as
\begin{equation}
\psi_+=\psi_1+\psi_2=\begin{pmatrix}
\hat{\Delta}\xi_1+(1-\hat{d}_0)\xi_2\\
-\hat{\Delta}\xi_2-(1+\hat{d}_0)\xi_1
\end{pmatrix},
\end{equation}
and find 
\begin{equation}
\psi_+^\dagger \psi_+=4+2(\xi^\dagger_1\hat{\Delta}\xi_2+\xi^\dagger_2\hat{\Delta}\xi_1). \label{psi-plus}
\end{equation}
The norm of $\psi_+$ is well defined if the matrix elements of $\hat{\Delta}$ in the above equation is zero. We observe that if $\xi_1=\chi_i$ and $\xi_2=\chi_j$ in Eq.~\eqref{psi-plus}, $\xi^\dagger_1\hat{\Delta}\xi_2$ is just the $(i,j)$-entry of $\Delta$. Since $\hat{\Delta}$ exactly has four zero entries for $\hat{\Delta}$, we can certainly construct a complete set of global wave functions, which are given by
\begin{equation}
\begin{split}
|1,\k\rangle &=\frac{1}{2}[\psi_1(\chi_2)-\psi_2(\chi_1)],\\
|2,\k\rangle &=\frac{1}{2}[\psi_1(\chi_1)+\psi_2(\chi_2)],\\
|3,\k\rangle &=\frac{1}{2}[-\psi_1(\chi_3)+\psi_2(\chi_4)],\\
|4,\k\rangle &=\frac{1}{2}[-\psi_1(\chi_4)-\psi_2(\chi_3)].
\end{split}
\end{equation}
The signs in the combinations above have been carefully chosen, such that the four wave functions are orthogonal. 
Explicitly, the orthonormal wave functions are
\begin{equation}
|1,\k\rangle=\frac{1}{2}\begin{pmatrix}
-1+\hat{d}_0\\
\hat{d}_3\\
\hat{d}_1\\
-\hat{d}_2\\
\hat{d}_3\\
-1-\hat{d}_0\\
\hat{d}_2\\
\hat{d}_1
\end{pmatrix},\quad 
|2,\k\rangle=\frac{1}{2}\begin{pmatrix}
\hat{d}_3\\
1-\hat{d}_0\\
\hat{d}_2\\
\hat{d}_1\\
-1-\hat{d}_0\\
-\hat{d}_3\\
-\hat{d}_1\\
\hat{d}_2
\end{pmatrix},\quad 
|3,\k\rangle=\frac{1}{2}\begin{pmatrix}
-\hat{d}_2\\
-\hat{d}_1\\
\hat{d}_3\\
1-\hat{d}_0\\
-\hat{d}_1\\
\hat{d}_2\\
1+\hat{d}_0\\
\hat{d}_3
\end{pmatrix},\quad
|4,\k\rangle=\frac{1}{2}\begin{pmatrix}
-\hat{d}_1\\
\hat{d}_2\\
-1+\hat{d}_0\\
\hat{d}_3\\
\hat{d}_2\\
\hat{d}_1\\
-\hat{d}_3\\
1+\hat{d}_0
\end{pmatrix}.
\end{equation}

Actually the wave functions are carefully ordered, and hence the Berry connection can be written in the concise form,
\begin{equation}
\mathcal{A}_{\alpha\beta,j}=\frac{1}{2}(\d_\alpha\partial_j\d_\beta-\d_\beta\partial_j \d_\alpha-\epsilon_{\alpha\beta\gamma\delta}\d_\gamma\partial_j\d_{\delta}).
\end{equation}
Substituting the above expression into the formula of the Chern-Simons form, we find
\begin{equation}
\begin{split}
Q_3(\A)&=-\frac{1}{8\pi^2}\epsilon^{ijk}\mathrm{tr}(\A_i\partial_j\A_k+\frac{2}{3}\A_i\A_j\A_k)\\
&=-\frac{1}{12\pi^2}\epsilon^{ijk}\epsilon^{\alpha\beta\gamma\delta}\d_\alpha\partial_i \d_\beta \partial_j \d_\gamma \partial_k \d_\delta,
\end{split}
\end{equation}
which is just the winding number of the unit vector valued in the unit sphere $S^3$ over the 3D BZ.

\subsection{The topological invariant }
Since the second Chern number is zero, the topological invariant is given by the boundary Chern-Simons terms, more clearly the difference of winding numbers $\mathcal{N}^\pi_W$ and $\mathcal{N}^0_W$ of $\hat{d}_\mu$'s in the two boundaries,
\begin{equation}
\mathcal{N}=\mathcal{N}^\pi_W-\mathcal{N}^0_W \mod 2.
\end{equation}
In the particular model, the winding numbers are determined by
\begin{equation}
\tilde{m}_\pi=m+1,\quad \tilde{m}_0=m-1,
\end{equation}
for the two boundaries, respectively. The results for typical masses in gapped regions of $m$ are summarized in the Tab.~\ref{Topological-Invariant}.
\begin{table}
	\begin{tabular}{c|c|c|c}
		$m$ & $\tilde{m}_\pi$ & $\tilde{m}_0$ & $\mathcal{N}=\mathcal{N}^\pi_W-\mathcal{N}^0_W$\\
		\hline
		5 & 6 &  4 & 0-0=0\\
		3 & 4  & 2 & 0-1=-1\\
		1 & 2  & 0 & 1+2=3\\
		-1 & 0 & -2 & -2+1=-1\\
		-3 & -2 & -4 & -1-0=-1\\
		-5 & -4 & -6 & 0-0=0		
	\end{tabular}
	\caption{The topological invariant for different $m$'s.\label{Topological-Invariant}}
\end{table}
We find that all gapped regions with $|m|<5$ correspond to the topologically nontrivial phase.

\section{Simulation with Ultracold Atoms}

For the simplicity of the simulation, we choose another set of Clifford algebra generators
\begin{equation}
\begin{split}
&\Gamma_0'=\sigma_3\otimes\sigma_0\otimes\sigma_0,\ \Gamma_1'=\sigma_1\otimes\sigma_1\otimes\sigma_1,\ \Gamma_2'=\sigma_1\otimes\sigma_1\otimes\sigma_3,\ \Gamma_3'=\sigma_1\otimes\sigma_3\otimes\sigma_0,\\
&\Gamma_4'=\sigma_2\otimes\sigma_0\otimes\sigma_0,\ \Gamma_5'=\sigma_1\otimes\sigma_2\otimes\sigma_0,\ \Gamma_6'=\sigma_1\otimes\sigma_1\otimes\sigma_2,
\end{split}
\end{equation}
which is related to the original ones by a unitary transformation that does not change the symmetry operator $\mathcal{RT}=\hat{\mathcal{K}}$. According, the Hamiltonian is transformed to be
\begin{equation}
\label{hh1}
\begin{split}
\H'(\k,k_{w})=&\sin k_w\Gamma_4'-\cos k_w\Gamma_0'+(m-\sum_{i=1}^{3}\cos k_i)\Gamma_0'\\
&+\sin k_1\Gamma_1'+\sin k_2\Gamma_2'+\sin k_3\Gamma_3'\\
&+t_0 \left(i\Gamma_3'\Gamma_5'\right),
\end{split}
\end{equation}
which has been divided into different parts for simulation.

\begin{figure}[h]
	\centering
	\includegraphics[scale=0.6]{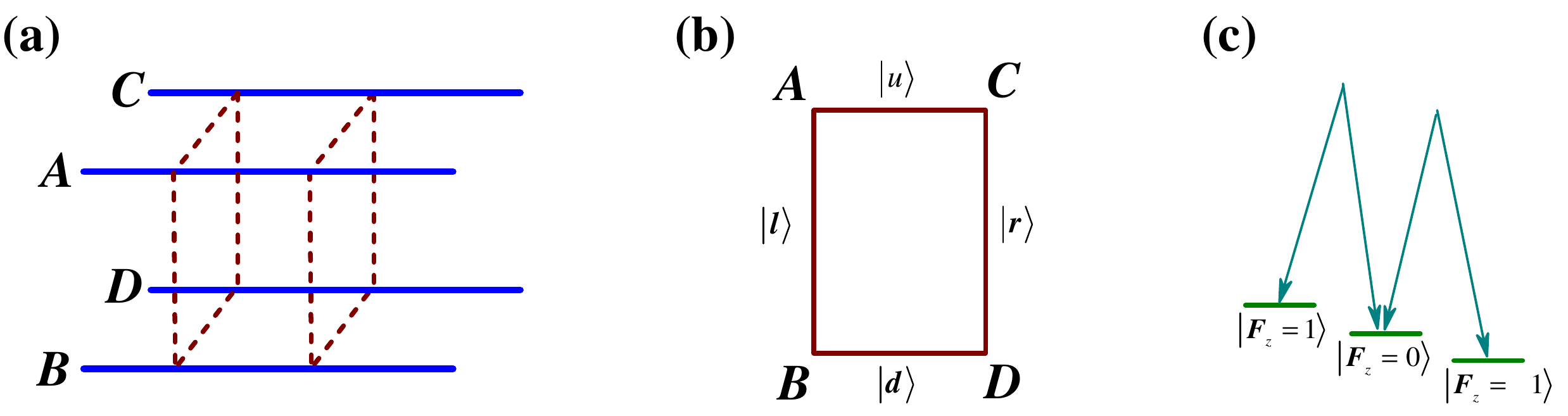}
	\caption{(a) The quasi 1D tetragonal optical lattice. The four parallel chains are denoted as $ A,B,C,D $. (b) The unit cell. (c) The internal ground levels of $ F=1 $ manifold split by an external magnetic field. \label{simulation}}
\end{figure}

We consider ultra-cold atoms with internal pseudospin moving in two ladders of optical lattice consisting of four chains denoted as $ A,B,C,D $ in Fig.~\ref{simulation}(a) to simulate the Hamiltonian in Eq.~\eqref{hh1}. The first two Pauli matrices of $ \Gamma' $ matrices act in the spaces spanned by $ \{\ket{l}, \ket{r}\} $ and $ \{\ket{u}, \ket{d}\} $, respectively. Then, $ \ket{\psi_{A}}=\ket{l}\otimes\ket{u} $, $ \ket{\psi_{B}}=\ket{l}\otimes\ket{d} $, $ \ket{\psi_{C}}=\ket{r}\otimes\ket{u} $, $ \ket{\psi_{D}}=\ket{r}\otimes\ket{d} $ in Fig.~\ref{simulation}(b).

The internal pseudospin is simulated by two levels $ \ket{F=1,F_{z}=-1} $ and $\ket{F=1,F_{z}=0}$ among the three levels with $ {F=1}$, which are slightly split by an external magnetic field, as shown in Fig.~\ref{simulation}(c). We set the amplitude of inter-chain nearest-neighbor (NN) hoppings as unit, and all next-nearest-neighbor (NNN) hoppings are introduced by the photon-assisted tunneling. When two counter-propagating Raman lasers along $ \mathbf{r} $ are resonant with the internal levels, the two-photon process can lead to the effective spin-orbital interaction,
\begin{equation}
H_{SO}=\left(\begin{array}{cc}
\frac{\hat{p}^{2}}{2M}+\delta & \Omega e^{2ik_{0}r} \\ \Omega e^{-2ik_{0}r} & \frac{\hat{p}^{2}}{2M}-\delta
\end{array}\right)\label{hso}
\end{equation}
in the subspace spanned by $ \{\ket{F=1,F_z=0}, \ket{F=1,F_z=-1}\} $. Here, $ \hat{p} $ is the momentum operator along $\mathbf{r}$, $\delta$ is the energy splitting by the Zeeman effect, $ k_{0} $ is the wave vector of the lasers along $\mathbf{r}$ and $ \Omega $ is the Rabbi frequency. To obtain a tight-binding model, the hopping coefficient $ T_{RR'} $ between sites $ \textbf{R} $ and $ \textbf{R}' $  can be derived as
\begin{align}
T_{RR'}=\bra{W_{\textbf{R}}}H_{SO}\ket{W_{\textbf{R}'}}, \label{tunnel}
\end{align}
where $ \ket{W_{\textbf{R}}} $ are the Wannier wave functions of the optical lattice.
The general form of $ T_{\textbf{R}\textbf{R}'} $ is $ t_0\sigma_0+t_1\sigma_1+t_2\sigma_2+t_3\sigma_3 $, with all $ t_i $ highly tunable. In addition, we note that all pseudospin-independent phases of hoppings can be tuned by the usual synthetic gauge field in cold-atom experiments.

\begin{figure}[h]
	\centering
	\includegraphics[scale=1]{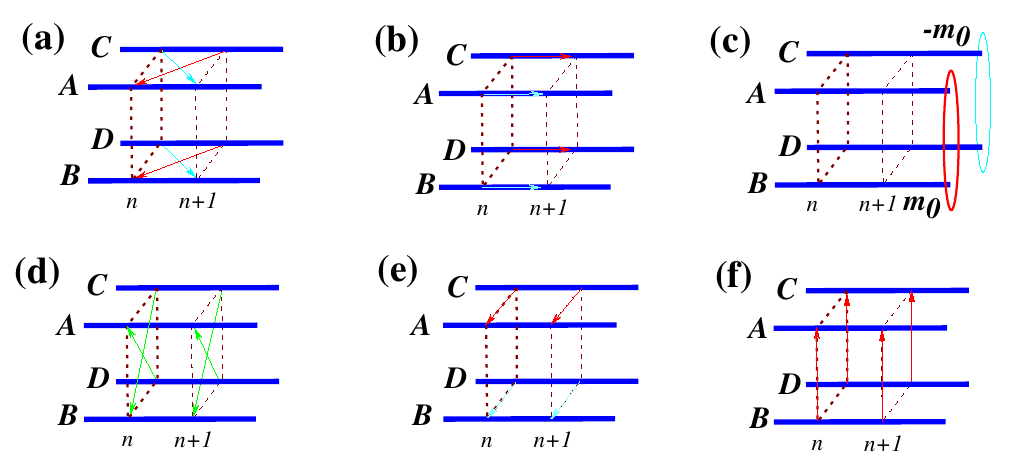}
	\caption{The simulation of all terms in Eq.~\eqref{hh1}. The light-blue and red arrows represent pseudospin-independent hoppings, and the former has an additional phase $ \pi $. The hoppings depicted by the green arrows involve pseudospin-orbital couplings.% (a) $ \sin k_{w}\Gamma_{4}' $. (b) $ -\cos k_{w}\Gamma_{0}' $. (c) $ \left(m-\sum_{j=1}^{3}\cos k_{j}\right)\Gamma_{0}' $. (d) $ \sum_{j=1,2} \sin k_{j}\Gamma_{j}'$. (e) $ \sin k_{3}\Gamma_{3}' $. (f) Partial mass term $ H_{1}=it\Gamma_{3}'\Gamma_{5}' $. 
		\label{hoppings}}
\end{figure}

\subsection{$ \sin k_{w}\Gamma_{4}' $}

This term can be translated to the tight-binding form as \begin{equation}
\begin{split}
&\sin k_{w}\Gamma_{6}=\sin k_{w} \sigma_{2}\otimes \sigma_{0}\otimes\sigma_{0}\\
\Rightarrow&\frac{1}{2}\sum_{n}\sum_{s=\uparrow,\downarrow}\Big[\left(f_{A,n,s}^{\dagger} f_{C,n+1,s}-f_{A,n+1,s}^{\dagger} f_{C,n,s}\right) + \left(f_{B,n,s}^{\dagger} f_{D,n+1,s}-f_{B,n+1,s}^{\dagger} f_{D,n,s}\right) \Big] +\mathrm{H.c.},
\end{split}
\end{equation}
where the first term corresponds to the NNN hoppings between two up-chains and the second term is the NNN hopping between two down-chains. Here and below, $ f_{i,n,s} $ with $ i=A,B,C,D $ is the annihilation operator of site $ n $ of the $ i $-chain with pseudospin $ s $. It is depicted in Fig.~\ref{hoppings}(a), where the light-blue arrows represent pseudospin-independent hoppings with an additional  phase $ \pi $, and the red arrows represent pseudospin-independent hopping without additional phase.

\subsection{$ -\cos k_{w}\Gamma_{0}' $}

This term only contains the NN inter-chain hoppings. There is an phase $ \pi $ for hoppings between two left-chains $ A,B $, and no phase for those between two right chains, $C$ and $D$. We then have
\begin{equation}
\begin{split}
&-\cos k_{w}\Gamma_{4}=-\cos k_{w} \sigma_{3}\otimes\sigma_{0}\otimes\sigma_{0}\\
\Rightarrow& -\frac{1}{2}\sum_{n}\sum_{s=\uparrow,\downarrow}\left(f_{A,n+1,s}^{\dagger} f_{A,n,s} +f_{B,n+1,s}^{\dagger} f_{B,n,s}\right) +\frac{1}{2}\sum_{n,s}\left(f_{C,n+1,s}^{\dagger} f_{C,n,s} +f_{D,n+1,s}^{\dagger} f_{D,n,s}\right)+\mathrm{H.c.},
\end{split}
\end{equation}
which is illustrated in Fig.~\ref{hoppings}(b).

\subsection{$ \left(m-\sum_{j=1}^{3}\cos k_{j}\right)\Gamma_0' $}

We regard $ m_{0}=m-\sum_{j=1}^{3}\cos k_{j} $ as an internal parameter of the cold-atom system. Then,
\begin{equation}
\begin{split}
&\left(m-\sum_{j=1}^{3}\cos k_{j}\right)\Gamma_{4}=m_{0}\sigma_{3}\otimes\sigma_{0}\otimes\sigma_{0}\\
\Rightarrow & m_{0}\sum_{n}\sum_{s=\uparrow,\downarrow}~\left(f_{A,n,s}^{\dagger} f_{A,n,s}+f_{B,n,s}^{\dagger} f_{B,n,s}\right)-\left(f_{C,n,s}^{\dagger} f_{C,n,s}+f_{D,n,s}^{\dagger} f_{D,n,s}\right),
\end{split}
\end{equation}
which means the on-site energy of a cold atom at two left-chains $ A,B $ is $ m_0 $, and that at two right-chains $ C,D $ is $ -m_0 $, as illustrated in Fig.~\ref{hoppings}(c). This term can be readily simulated by applying a tilted $ W $-potential.

\subsection{$ \sum_{j=1,2} \sin k_{j}\Gamma_{j}'$}
Defining $ t_{j}=\sin k_{j} $, for $ j=1,2 $, this term gives the NNN pseudospin-dependent hoppings by 
\begin{equation}
\begin{split}
&\sum_{j=1,2} \sin k_{j}\Gamma_{j}'=\sigma_{1}\otimes\sigma_{1}\otimes\left(t_{1}\sigma_{1}+t_{2}\sigma_{3}\right)\\
\Rightarrow& t_1\sum_{n}\Big[\left(f_{A,n,\uparrow}^{\dagger}f_{D,n,\downarrow}+f_{A,n,\downarrow}^{\dagger}f_{D,n,\uparrow}\right)+\left(f_{B,n,\uparrow}^{\dagger}f_{C,n,\downarrow}+f_{B,n,\downarrow}^{\dagger}f_{C,n,\uparrow}\right)\Big]+\mathrm{H.c.}\\
&+t_2\sum_{n}\Big[\left(f_{A,n,\uparrow}^{\dagger}f_{D,n,\uparrow}-f_{A,n,\downarrow}^{\dagger}f_{D,n,\downarrow}\right)+\left(f_{B,n,\uparrow}^{\dagger}f_{C,n,\uparrow}-f_{B,n,\downarrow}^{\dagger}f_{C,n,\downarrow}\right)\Big]+\mathrm{H.c.},
\end{split}
\end{equation}
as illustrated in Fig.~\ref{hoppings}(d). The pseudospin is flipped during the hoppings of term $ t_{1}\sigma_{1}\otimes\sigma_{1}\otimes\sigma_{1} $. For term $ t_{2}\sigma_{1}\otimes\sigma_{1}\otimes\sigma_{3} $, an additional phase $ \pi $ is acquired for pseudospin $ \ket{\downarrow} $ during the hoppings. Here, the photon-assisted tunneling is applied to realize the hoppings between chain $ A $ ($ B $) and chain $ D $ ($ C $), according to Eqs.~\eqref{hso} and \eqref{tunnel}.

\subsection{$ \sin k_{3}\Gamma_{3}' $}
Let $ t_{3}=\sin k_{3} $, we have
\begin{equation}
\begin{split}
&\sin k_{3}\Gamma_{3}=t_{3} \sigma_{1} \otimes \sigma_{3} \otimes\sigma_{0}\\
\Rightarrow& t_3\sum_{n}\sum_{s=\uparrow,\downarrow}\left(f_{A,n,s}^{\dagger}f_{C,n,s}-f_{B,n,s}^{\dagger}f_{D,n,s}\right)+\mathrm{H.c.},
\end{split}
\end{equation}
which represents NN hopping between two up (down) chains in Fig.~\ref{hoppings}(e). An additional phase $ \pi $ is required for the hopping between two down chains $ B,D $. This $ \pi $ phase can be induced by rotating the optical lattice along $ w $ direction to obtain an effective gauge field by the Coriolis effect.  

\subsection{Partial Mass Term $ H_{1}=it\Gamma_{3}'\Gamma_{5}' $}

The tight-binding model for this term is 
\begin{align}
&it\Gamma_{3}\Gamma_{7}=t \sigma_{0}\otimes\sigma_{1}\otimes \sigma_{0}\nonumber\\
\Rightarrow& t\sum_{n}\sum_{s=\uparrow,\downarrow}\left(f_{A,n,s}^{\dagger}f_{B,n,s}+f_{D,n,s}^{\dagger}f_{C,n,s}\right)+H.c.
\end{align}
which are just NN hoppings shown by red arrows in Fig.~\ref{hoppings}(f) between two left (right) chains.

\end{document}